\documentclass[12pt]{article}
\usepackage{amsmath,mathrsfs,bm,amssymb,color,theorem,hiroshima}
\usepackage{tikz-cd}
\usetikzlibrary{calc} % 
\usepackage{latexsym}
\usepackage{graphicx}
\usepackage{subcaption} %
\newtheorem{theorem}{Theorem}[section]
\newtheorem{proposition}[theorem]{Proposition}
\newtheorem{lemma}[theorem]{Lemma}
\newtheorem{corollary}[theorem]{Corollary}
\newtheorem{definition}[theorem]{Definition}
\newtheorem{example}[theorem]{Example}
\newtheorem{remark}[theorem]{Remark}

\makeatletter
\@addtoreset{equation}{section}
\makeatother

\begin{document}
\title{Representations of Josephson junction on the unit circle and the derivations of 
Mathieu operators and Fraunhofer patterns}
\author{Toshiyuki Fujii\footnote{Asahikawa Medical University, Department of Physics}, Fumio Hiroshima\footnote{Faculty of Mathematics, Kyushu University} and 
Satoshi Tanda\footnote{Faculty of Engeneering, Hokkaido University}}
%\date{\today}
\maketitle

\begin{abstract}
The Hamiltonian $\jj$ of the Josephson junction is introduced as a self-adjoint operator on $\ln \otimes \ln$.
It is shown that $\jj$ can also be realized as a self-adjoint operator $H_{S^1}$ on 
$L^2(S^1)\otimes L^2(S^1)$, from which a Mathieu operator 
%$-\frac{d^2}{d\theta ^2}-2\alpha\cos \theta$, 
is derived.
A fiber decomposition of $H_{S^1}$ with respect to the total particle number is established, and the action on each fiber is analyzed.
In the presence of a magnetic field,  a phase shifts defines the magnetic Josephson junction Hamiltonian 
$H_{S^1}(\Phi)$ and the Josephson current $I_{S^1}(\Phi)$.
For a constant magnetic field inducing a local  phase shift  $\Phi(x)$, the corresponding local current 
$I_{S^1}(\Phi(x))$ is computed, and it is proved that the Fraunhofer pattern arises naturally.
\end{abstract}
\newpage{\tableofcontents}\newpage
\section{Introduction}
The Josephson junction \cite{jos62} is a fundamental component in superconducting circuits and it is characterized by the coherent tunneling of Cooper pairs between two superconductors separated by a thin insulating barrier. 
This quantum mechanical phenomenon gives rise to rich physical behavior, including persistent supercurrents and quantized voltage steps.  
Mathematically, 
two superconductors are described by Hilbert spaces $\cH_A$ and $\cH_B$, 
and the total system of the Josephson junction is given by 
$\cH=\cH_A\otimes \cH_B$. 
More precisely $\cH_A=\cH_B=\ln$. Then 
\begin{align*}\cH=\ln\otimes\ln.\end{align*} 
Let 
$S^1=\{e^{i\theta}\mid \theta\in[0,2\pi)\}$ 
be 
the unit circle. 
and 
$\rd\mu(\theta)={\rd \theta}/{2\pi}$ the normalized Lebesgue measure on $S^1$.  
$L^2(S^1)$ is the Hilbert space
\[
L^2(S^1)=\lkk f:S^1\to\CC \Big|\ 
   \int_{S^1}|f(z)|^2\rd\mu(z)<\infty \rkk,
\]
with inner product
$
( f,g)_{L^2(S^1)}=\int_{S^1}\overline{f(z)}g(z)\rd\mu(z)$. 
Since $e^{i\theta}=e^{i(\theta+2\pi)}$, functions in $L^2(S^1)$ are automatically $2\pi$-periodic.  
Thus $L^2(S^1)$ can be identified with the subspace of $L^2([0,2\pi))$ consisting of $2\pi$-periodic functions.  
We shall use this identification without further mention. 
The Josephson junction can be modeled by a quantum system in which the phase across the junction is a $2\pi$-periodic variable. 
This periodicity naturally leads to a formulation on the Hilbert space 
\begin{align*}\cH_{S^1}=L^2(S^1)\otimes L^2(S^1)\end{align*} 
instead of 
$\ln\otimes \ln$. 
In this setting, the phase operator $\theta$ acts as a multiplication operator, while the conjugate charge operator is realized as a $(-i)$ times differential operator, i.e., $-i\frac{d}{d\theta}$. 
These two operators {\it formally} satisfy the canonical commutation relation $[-i\frac{d}{d\theta},\theta]=-i\one$, 
but due to the compactness of the unit circle $S^1$, 
a careful functional analytic treatment is required, 
since the multiplication by $\theta$ is not periodic.

In physical literatures the Hamiltonian of the Josephson junction typically takes the form
\begin{align}\label{JJ}
H = {4E_C} \left(-i\frac{d}{d\theta}\right)^2-E_J \cos \theta,
\end{align}
where $E_C=e^2/2C$ is the charging energy with charge $e$ and junction capacitance $C$, 
and $E_J$ is the Josephson coupling constant. 
The potential $-E_J \cos \theta$ reflects the tunneling of Cooper pairs. 
This potential is $2\pi$-periodic and corresponds to a potential defined on the circle $S^1$. 
In this paper \eqref{JJ} is referred to as the Mathieu operator \cite{mcl47}.  
When a constant magnetic field $B=(0,0,b)$ is applied to a Josephson junction of width 
$W=1$ and 
the barrier thickness  $d$, 
the phase difference varies linearly across the junction :
$\theta (x) = \theta + \frac{2\pi}{\Psi_0} \Psi x$ 
with $\Psi_0 = 1/2e$, 
where $\Psi=bd$ denotes the magnetic flux.  
The total current is obtained as the superposition of local Josephson currents:
\begin{align}
\label{fh}
 j_c \int_{-1/2}^{1/2} \sin(\theta (x))  \rd x
= j_c  \frac{\sin(\pi \Psi/\Psi_0)}{\pi \Psi/\Psi_0} \sin \theta,
\end{align}
where $j_c$ denotes the critical current density of the junction. 
This sinc-like dependence of the total current on the magnetic flux $\Psi$ 
is known as the {Fraunhofer pattern}, directly analogous to the 
single-slit diffraction pattern in optics.

There exists a huge number of works on the derivation of Mathieu operators 
and Fraunhofer patterns from Josephson junction models. 
In \cite{ADS01}, the Mathieu operator is obtained from the two-particle Bose-Hubbard model, albeit by invoking the Dirac phase operator. A related discussion also appears in 
\cite{Uchino2021, ban91,ban95,HTKT00}; however, the treatment there remains largely heuristic and falls short of a fully rigorous mathematical formulation. 
In \cite{Petersen2013}, the Josephson junction Hamiltonian is analyzed as a self-adjoint operator in the setting of a cavity system, while in \cite{Bacsi2025} the effective Hamiltonian is derived from BCS theory through a Schrieffer-Wolff transformation, incorporating quasiparticle effects. 
Earlier works such as \cite{Giordano2015} address the transition from microscopic to macroscopic descriptions. 
%In \cite{FAN200313} a path integral approach is also discussed. 
%
On the other hand the Fraunhofer pattern in Josephson junctions is typically derived by integrating the local current $I_{\mathrm{JJ}}(\Phi(x))$ across the junction width under a constant magnetic field, which induces a linear phase with respect to $x$. 
Departures from the ideal Fraunhofer pattern have also been studied in various settings, such as diffusive junctions \cite{Montambaux2007} and magnetic barriers \cite{Borcsok2019}.

Although these contributions provide valuable insights, they remain far from firm mathematical rigor, being mainly heuristic, intuitive, or discovery-oriented in nature. 
Without mathematical rigor, treatments of the Josephson junction Hamiltonian suffer from unclear operator domains, lack of self-adjointness, possible misinterpretations of the spectral structure, and the use of intuitive approximations that may lead to further inconsistencies. 
Heuristic or intuitive approaches obscure the precise conditions under which phenomena such as the Mathieu operator 
and the Fraunhofer pattern arise, and hinder systematic extensions of the theory to more general settings. 
This highlights the necessity of a fully rigorous operator-theoretic formulation based on the theory of Hilbert spaces. 
To the best of our knowledge, no prior study has succeeded in deriving, in a mathematically rigorous manner, 
either the Mathieu operator or the Fraunhofer pattern starting directly from the Josephson junction 
Hamiltonian defined on $\ln \otimes \ln $. 
Given the remarkable progress in the mathematical foundations of quantum mechanics and quantum field theory in recent decades, it is striking that a comparable level of rigor has not yet been fully realized in the study of the Josephson junction. 
In this paper, we aim to close that gap, providing for the first time a mathematically precise derivation that unites the Josephson junction with the Mathieu operator and the Fraunhofer pattern. 
In doing so, the paper not only establishes a new bridge between physics and mathematics but also elevates the study of 
Josephson systems into the realm of rigorous mathematical analysis.
Henceforth, we abbreviate \lq\lq Josephson junction" as~JJ.

%Let \begin{align*}\cH_{S^1}=L^2(S^1)\otimes L^2(S^1).\end{align*}
In this paper, we develop a concrete realization of the JJ-Hamiltonian 
$\jj $ on 
$\cH_{S^1}$, starting from its definition on 
$\ln \otimes \ln $.
$\jj $ is of the form:
\begin{align*}\jj =\frac{1}{2C}(\sM\otimes\one-\one\otimes \sM -q)^2-\alpha (L\otimes L^\ast+L^\ast\otimes L),\end{align*}
where 
$\sM$ is the number operator on $\ln$, $L$ is a unilateral shift operator on $\ln$, 
and 
$q,C,\alpha\in\RR$ are constants. 
$\sN=\sM\otimes\one-\one\otimes \sM $ denotes the relative number operator, 
and $ H_T=L\otimes L^\ast+L^\ast\otimes L$ describes a tunneling process. 
See \eqref{D1} for the definition of $\jj $. 
Here we set $e=1$,  the constant $q$ serves as a gauge shift and $\alpha$ a coupling constant corresponding to $E_J$ of \kak{JJ}. 
It commutes the total number operator $\tN=\sM\otimes \one+\one\otimes \sM$:  
\begin{align*}[\tN,\mm]=0,\end{align*} and hence $\jj$ can be reduced to the $k$-particle subspace of $\ln\times\ln$ for any $k\geq0$. 
We construct a seriese of unitary operators $S_f,u,\rho,U$ and $\sF $ 
such that 
\begin{align*}
\cH=\ln\otimes\ln\stackrel{S_f}{\longrightarrow}\ZN
\stackrel{u}{\longrightarrow} \lz\otimes\ln
\stackrel{\rho}{\longrightarrow} (\lz\oplus\lz)\otimes\ln 
\stackrel{U}{\longrightarrow} \lz\otimes\lz 
\stackrel{\sF }{\longrightarrow} \cH_{S^1}.
\end{align*}
By virtue of inner automorphisms $T_V v=V v V^{-1}$ induced 
by 
these unitary operators, the JJ-Hamiltonian $\jj $ is transformed as 
\begin{align*}
\jj \stackrel{T_{S_f}}{\longrightarrow}\hh^{f}
\stackrel{T_u}{\longrightarrow} \hh^{u} 
\stackrel{T_\rho}{\longrightarrow} \hh^{\rho}
\stackrel{T_U}{\longrightarrow} \hh^{U}
\stackrel{T_{\sF }}{\longrightarrow} H_{S^1}.
\end{align*}
See \eqref{D2} for the definition of $\hh^{f}$, 
\eqref{D3} for that of $\hh^{u} $, 
\eqref{D5} for that of $\hh^{\rho}$, 
and \eqref{D6} for that of $\hh^{U}$.
Finally, we construct a unitary operator $\mathcal{U}$ obtained as the composition of these unitaries: 
\begin{align*}
 \cU:\ \cH\ \longrightarrow\ \cH_{S^1},
 \quad H_{S^1} = \cU \jj \cU^{-1},
\end{align*}
so that $H_{S^1} $ provides the desired representation like \eqref{JJ}. 
%See Figure \ref{uu}. 
Under this identification, the relative number operator $\sN $ is carried to the first order differential operator
$- i\frac{\partial}{\partial\theta_1}$ on the appropriate circle variable, while the operator $H_T$ becomes multiplication by $e^{\pm i\theta_1}$ and 
$e^{\pm i\theta_2}$
on the circle coordinate $(\theta_1,\theta_2)$ compositing with projections. 
We arrive at the model of the form
\begin{align*}
H_{S^1}=
\frac{1}{2C}
\lk 
-2i\frac{\partial}{\partial \theta_1}+q
\rk^2\otimes P _{[0,\infty)}
+
\frac{1}{2C}
\lk 
-2i\frac{\partial}{\partial \theta_1}+\one+q
\rk^2\otimes P _{(-\infty,-1]}
-\alpha H_{S^1,T}
\end{align*}
on $\cH_{S^1}$.
See \eqref{D4} for the definition of $H_{S^1}$. 
We verify that $H_{S^1} $ is self-adjoint on the natural Sobolev domain inherited from $D(\sN ^2)$ and bounded from below. 
Restricting 
$H_{S^1}$ we derive the Mathieu operator \eqref{JJ}. 
Furthermore we shall discuss the Josephson current. 
The magnetic Hamiltonian of the Josephson junction is defined by 
\begin{align*}\mm =\frac{1}{2C}(\sM\otimes\one-\one\otimes \sM -q)^2-\alpha (e^{i\Phi}L\otimes L^\ast+e^{-i\Phi}L^\ast\otimes L)\end{align*}
for $\Phi\in\RR$. Here $\Phi$ describes the phase shift. 
We see that 
$[\sN,\mm]\not=0$, and the Josephson current is defined by 
\begin{align*}I_{\rm JJ}(\Phi)=i\half [\sN,\mm].\end{align*} 
We shall develop a rigorous mathematical derivation of how the current is altered under the influence of the  phase shift  $\Phi\in\RR$. 
Let $I_{S^1}(\Phi)$ be the representation of $I_{\rm JJ}(\Phi)$ on $\cH_{S^1}$. 
We rigorously prove that, under a constant magnetic field, the local  phase shift  
$\Phi(x)$, $-1/2 \leq x \leq 1/2$, 
across the Josephson junction leads to a total Josephson current from which the Fraunhofer pattern emerges:
\begin{align}\label{fr}
\int_{-1/2}^{1/2} 
(\psi, I_{S^1}(\Phi(x))\psi)_{\cH_{S^1}} \rd x
= \frac{\sin(\Psi/2)}{\Psi/2} (\psi, I_{S^1}(\Phi(0))\psi)_{\cH_{S^1}}.
\end{align}

This paper is organized as follows.
In Section~\ref{s2}, we introduce the total Hamiltonian $ \jj $ on $ \mathcal{H}$.
In Section~\ref{ss}, we define the total Hamiltonian $ \hh^{f}$ on $ \ZN$.
In Section~\ref{s3}, we define the total Hamiltonians $ \hh^{u}$ on $ \lz \otimes \ln$ and $ \hh^{\rho}$ on $(\lz \oplus \lz) \otimes \ln$.
We also construct the unitary operator $U$ implementing the equivalence between $(\lz \oplus \lz) \otimes \ln$ and $ \lz \otimes \lz$, and provide the explicit form of $ \hh^{U}$.
In Section~\ref{s4}, we state the main theorem (Theorem~\ref{main}). 
% and derive the Mathieu operators \eqref{JJ} in Remark~\ref{ma}.
In Section~\ref{s5}, we discuss the fiber decomposition of the JJ-Hamiltonian and also derive the Mathieu operators on each fiber in \eqref{ma1}.
Section~\ref{7} is devoted to an investigation of the Josephson current of the magnetic JJ-Hamiltonian, where we derive the Fraunhofer pattern \eqref{fr} and reveal the emergence of the Aharonov-Bohm effect from the Josephson current.
Finally, in Section~\ref{8}, we discuss an array of $n$ junctions as concluding remarks.

\section{JJ-Hamiltonian and magnetic JJ-Hamiltonian}
\label{s2}
\subsection{JJ-Hamiltonian $\jj$}
%\subsection{Relative number operators and tunneling processes}
We denote the set of natural numbers by $\NN=\{0,1,2,\ldots\}$. 
Let 
\begin{align*}
\ln =
 \Bigl\{ a = (a_n)_{n \in \NN} \Big| \sum_{n \in \NN} |a_n|^2 < \infty \Bigr\}
\end{align*}
denote the Hilbert space of square-summable sequences, endowed with the inner product
$(a,b) = \sum_{n \in \NN} \overline{a_n} b_n$.
Note that the map $a \mapsto (a,b)$ is anti-linear, while $b \mapsto (a,b)$ is linear.
Let \begin{align*}\phi_n=(\delta_{mn})_m\in \ln ,\quad 
\delta_{mn}=\aaa {1&m=n,\\0&m\neq n.} \end{align*}
Then $\{\phi_n\}_{n\in\NN}$ is a complete orthonormal system of $\ln $. 
Let $\sM$ be the number operator in $\ln $. 
Then $\sM a=\sum_{n\in\NN} na_n\phi_n$ for $a=\sum_{n\in\NN} a_n\phi_n$ and the domain of 
$\sM$ is given by 
$$D(\sM)=\lkk a=(a_n)_{n\in\NN}\in\ln \left| \sum_{n\in\NN} |na_n|^2<\infty\right.\rkk.$$
In particular 
$\sM \phi_n=n\phi_n$ for any $n\in\NN$. 
Notation $\lz$ also denotes the set of square summable sequences on the integer $\ZZ$, 
and the number operator in $\lz$ is denoted by $N$ and its domain is given by 
$$D(N)=\lkk a=(a_n)_{n\in\ZZ}\in\lz \left| \sum_{n\in\ZZ} |na_n|^2<\infty\right.\rkk.$$
Now we define the total Hilbert space for the Josephson junction. 
Let $\cH_A=\ln $ and 
$\cH_B=\ln $.
The total Hilbert space of the Josephson junction is defined by %the tensor product of $\cH_A$ and $\cH_B$: 
\begin{align*}\cH=\ln \otimes \ln .\end{align*}
%Let $N_A$ and $N_B$ be the number operators in $\cH_A$ and $\cH_B$, respectively. 
We define the relative number operator $\sN$ by 
\begin{align*}\sN =\sM\otimes \one-\one\otimes \sM\end{align*} 
and the total number operator $\tN$ by 
\begin{align*}\tN =\sM\otimes \one+\one\otimes \sM.\end{align*} 
It follows that 
$\sN \phi_n\otimes\phi_m=(n-m)\phi_n\otimes\phi_m$ and 
$\tN \phi_n\otimes\phi_m=(n+m)\phi_n\otimes\phi_m$ for any $n,m\in\NN$.
Let $\s(T)$ denote the spectrum of $T$. 
Since $ \{\phi_n\otimes\phi_m\}_{(n,m)\in\NN\times \NN}$ is a complete orthonormal system of 
$\cH$, it can be seen that $\s(\sN)=\ZZ$ and 
the multiplicity of each $m\in \ZZ$ is infinity, 
while
 $\s(\tN)=\NN$ and 
the multiplicity of each $m\in \NN$ is $m+1$. 
From a physical standpoint, 
$\sN $
 represents the difference in the particle numbers associated with the subsystems 
$\cH_A$ and $\cH_B$. 
It thus provides a precise operator-theoretic manifestation of the particle number asymmetry between the two components of the quantum system.
The kinetic Hamiltonian is defined by 
\begin{align*} H_C=\frac{1}{2C}(\sN -q)^2.\end{align*}
\begin{lemma}
It follows that $\s( H_C)=\{\frac{1}{2C}(n-q)^2\}_{n\in\NN}$ and 
the multiplicity of each eigenvalue $\frac{1}{2C}(n-q)^2$ is infinity. 
\end{lemma}
\proof
Since 
$\s(\sN)=\ZZ$ and 
the multiplicity of each $m\in \ZZ$ is infinity, 
the lemma follows. 
\qed
Now let us define the tunneling Hamiltonian. 
Let $L:\ln \to \ln $ be 
the unilateral shift defined by 
\begin{align*}
&L\phi_{m}=\aaa{\phi_{m-1} &m\geq 1,\\
0& m=0,}\\
&L^\ast \phi_{m}=\phi_{m+1}.
\end{align*}
Therefore
 $LL^\ast=\one$ and $L^\ast L=\one- P _0$, where $ P _0$ denotes the projection onto 
$\overline{\rm LH}\{\phi_0\}$. Here $\overline{\rm LH}\cK$ denotes the closed linear hull of $\cK$. 
Moreover $[\sM, L]=-L$ and $[\sM,L^\ast]=L^\ast$ hold true on a dense domain.  
We consider that one particle transfers from $\cH_A$ to $\cH_B$, which is defined by
\begin{align}\label{t}
(L\otimes L^\ast) \phi_n\otimes \phi_m= \phi_{n-1}\otimes \phi_{m+1}\quad n\geq1, m\geq0.
\end{align}
%See Figure \ref{z} (a).
In a similar manner we consider that one particle transfers from $\cH_B$ to $\cH_A$, 
which is defined by
\begin{align}\label{tt}
(L^\ast \otimes L) \phi_n\otimes \phi_m= \phi_{n+1}\otimes \phi_{m-1}\quad n\geq0, m\geq1.
\end{align}
According to \eqref{t} and \eqref{tt} the tunneling Hamiltonian is defined by 
\begin{align*} H_T=L\otimes L^\ast+L^\ast\otimes L.\end{align*}

\begin{definition}[JJ-Hamiltonian]
The total Hamiltonian of the Josephson junction is defined by 
\begin{align}\label{D1}
\jj = H_C-\alpha H_T,
\end{align}
where $\alpha\in\RR$ is the coupling constant. 
\end{definition}

\subsection{Magnetic JJ-Hamiltonian $\mm$ and Josephson current}
We show a back ground of the  phase shift  $\Phi\in\RR$. 
Let us consider a magnetic field $B:\RR^3\to\RR^3$ and suppose that 
\begin{align*}
B = \nabla \times A,
\end{align*}
where $A:\RR^3\to\RR^3$ is a vector potential. 
The  phase shift  $\Phi$ due to the magnetic field $B$ is given by
\begin{align}\label{A}
\Phi = \int_{C_{\rm JJ}} A \cdot \rd r,
\end{align}
where $C_{\rm JJ}$ denotes the path across the junction barrier.

In what follows, we consider $\Phi$ to be a parameter ranging over $\RR$.
We define the magnetic tunneling Hamiltonian by 
\begin{align*} H_T(\Phi) =e^{i\Phi}L\otimes L^\ast+e^{-i\Phi}L^\ast\otimes L.\end{align*}
\begin{definition}[Magnetic JJ-Hamiltonian]
The magnetic JJ-Hamiltonian is defined by 
\begin{align}\label{MD1}
\mm = H_C-\alpha H_T(\Phi).
\end{align}
\end{definition}

\begin{lemma}
$\mm$ is self-adjoint on $D(\sN ^2)$ and essentially self-adjoint on any core of 
$\sN ^2$, and bounded from below 
for any $C,q, \alpha,\theta \in\RR$. 
Moreover 
\begin{align}
\label{com}
[\tN , \mm]=0.
\end{align} 
\end{lemma}
\proof
It can be seen that 
${H}_T(\Phi)$ is a self-adjoint bounded operator with 
$\lvert {H}_T(\Phi) \rvert \leq 2$. 
Since ${H}_T(\Phi)$ is bounded, it follows that 
$\mm$ is self-adjoint on $D((\sN+q)^2)$ and 
essentially self-adjoint on any core of 
$(\sN+q)^2$, and it is bounded from below 
for any $C, q, \alpha \in \RR$ by the Kato-Rellich theorem~\cite{kat51}. 
Since $D((\sN+q)^2) = D(N^2)$ for any $q \in \RR$, 
and since the cores of $(\sN+q)^2$ and $\sN^2$ coincide, the lemma follows. 
Moreover, since $[\sM,L] = -L$ and $[\sM,L^{\ast}] = L$, 
we obtain 
\begin{align*}
 [\tN, e^{i\Phi}L \otimes L^{\ast} + e^{-i\Phi}L^{\ast} \otimes L] = 0.
\end{align*} 
Together with $[\tN, H_C] = 0$, equation~\eqref{com} follows.
\qed
The introduction of the  phase shift  can be realized as a unitary transformation. 
This is stated in the following lemma. 
\begin{lemma}[Gauge transformation]
\label{gauge}
Let $\Phi\in\RR$. Then 
\begin{align*}e^{-i(\Phi/2)\sN} \jj e^{i(\Phi/2) \sN}=\mm.
\end{align*}
\end{lemma}
\proof
It is easy to see that 
$e^{-i\Phi\sM}Le^{i\Phi\sM}\phi =e^{i\Phi}L\phi$ and 
$e^{-i\Phi\sM}L^\ast e^{i\Phi\sM}\phi =e^{-i\Phi}L^\ast\phi$ 
for any $\phi\in \ln$,  and 
$e^{i\Phi\sN}=e^{i\Phi\sM}\otimes e^{-i\Phi\sM}$.
Combining these formulas we can see that 
\begin{align*}e^{-i\Phi\sN}(L\otimes L^\ast+L^\ast \otimes L)e^{i\Phi\sN}=
e^{2i\Phi}L\otimes L^\ast+e^{-2i\Phi}L^\ast \otimes L.\end{align*}
Moreover 
it can be seen that 
$e^{-i\Phi\sN} H_C e^{+i\Phi\sN}=H_C$ on 
$D(H_C)$. Then the lemma follows. 
\qed

\begin{example}[Constant magnetic field]\label{constant }
Consider a Josephson junction characterized by a barrier thickness $d$ and a width $W$. 
We adopt the Cartesian coordinate system $(x,y,z)$ such that the $x$-axis is parallel to the junction width, 
the $y$-axis is parallel to the barrier, and the $z$-axis is perpendicular to the junction. 
Consider a constant  magnetic field 
\begin{align*}
B(x,y,z) = (0,0,b),
\end{align*}
which can be expressed as
\begin{align*}
B = \nabla \times A, \qquad A(x,y,z) = (0,bx,0).
\end{align*}
The  phase shift  $\Phi$ induced by the magnetic field is given by \eqref{A}, 
where $C_{\rm JJ}$ denotes the path across the junction barrier in the $y$-direction: 
\begin{align*}
C_{\rm JJ}:\ r(t) = (x,t,0), \qquad -d/2 \leq t \leq d/2.
\end{align*}
Carrying out the integration, the  phase shift  at position $x$ is obtained as 
\begin{align*}
\Phi = \Phi(x) = b d x.
\end{align*}
Accordingly, for $-W/2 \leq x \leq W/2$, the  phase shift  varies linearly in $x$.
We shall discuss the magnetic JJ-Hamiltonian associated to a constant magnetic field in Section \ref{7}. 
\end{example}

It is shown above that 
$[\tN , \mm]=0$, whereas 
\begin{align}
\label{phi}
[N_-, \mm]=2\alpha(e^{i\Phi}L\otimes L^\ast-e^{-i\Phi}L^\ast\otimes L)\not=0.
\end{align}
The Josephson current is defined below. 
\begin{definition}[Josephson current]
The Josephson current is defined by 
\begin{align*}
I_{\rm JJ}(\Phi) = i\half [N_-, \mm].
\end{align*} 
\end{definition}
By \eqref{phi} it is expressed as 
\begin{align}\label{ttt}
I_{\rm JJ}(\Phi) = i\alpha(e^{i\Phi} L\otimes L^\ast-e^{-i\Phi} L^\ast\otimes L).
\end{align}
We are interested in the map $\Phi \mapsto (\psi, I_{\rm JJ}(\Phi)\psi)$ and we shall discuss this in Section \ref{7}. 
\begin{lemma}\label{bounded}
For all $\Phi\in\RR$, $I_{\rm JJ}(\Phi)$ is a bounded operator. 
\end{lemma}
\proof
By \kak{ttt} we see that 
$\|I_{\rm JJ}(\Phi)\|\leq |\alpha|$. 
Then the proof is complete. 
\qed

\section{JJ-Hamiltonian $\hh^f$ on $\ZN$}
\label{ss}
Hereafter, we will primarily discuss $\jj$ in place of $\mm$, for the sake of simplicity.
\subsection{Alternative complete orthonormal system $\{\Phi_{n,m}\}_{(n,m)\in\ZZ\times \NN}$ 
in $\N$}
Under the identification 
$a\otimes b\cong (a_nb_m)_{n,m\in\NN\times\NN}$, 
we can see that 
$\ln \otimes \ln \cong\lnn $.
Henceforth we study $\lnn $ instead of $\ln \otimes \ln $. 
The subset $\ZZ_m$ of $\NN\times\NN$ is defined by 
$$\ZZ_m=\{(m+n,m), (m,m), (m,m+n)\in \NN\times \NN\mid n\in \NN\}.$$ 
Then $\ZZ_m\cong \ZZ$ 
by the bijection 
$i_m:\ZZ_m\to\ZZ$, 
where 
\begin{align*}i_m :(m,m)\mapsto 0,\quad 
i_m:(m+n,m)\mapsto n,\quad i_m:(m,m+n)\mapsto -n.\end{align*} 
See Figure \ref{x}. 
\begin{figure}[t]
\begin{align*}
\begin{tikzpicture}[scale=0.8]
 \def\N{6} % 表示する最大の整数座標
 % 軸
 \draw[very thick, ->] (0,0) -- (\N+0.7,0) node[below] {$m$};
 \draw[very thick,->] (0,0) -- (0,\N+0.7) node[left] {$n$};
 % 第一象限のグリッド
 \draw[step=1cm,gray!40] (0,0) grid (\N,\N);
 % 格子点
 \foreach \i in {0,...,\N}{
 \foreach \j in {0,...,\N}{
 \fill (\i,\j) circle (1.6pt);
 }
}
 % 特殊な点
 \node[below right] at (3,3) {${}_{(m+1,m+1)}$};
 \node[below right] at (2,2) {${}_{(m,m)}$};
 \node[right] at (6,2) {$\ZZ_m\cong \ZZ$};
 \node[right] at (6,3) {$\ZZ_{m+1}\cong \ZZ$};
 \node[above right] at (6,6) {$\NN$};
 % 対角成分 m=n を太い黒線でつなぐ
 \draw[ultra thick, black] (0,0) -- (\N,\N);
 % 追加の線 (2,6)--(2,2), (2,2)--(6,2)
\draw[thick] (2,6) -- (2,2) -- (6,2);
\draw[thick] (3,6) -- (3,3) -- (6,3);
\end{tikzpicture}
\end{align*}
\caption{$\NN\times \NN \cong \ZZ\times \NN$ by $f=i_X\circ i$}
\label{x}
\end{figure}
Let 
$X=\bigcup_{m=0}^\infty \ZZ_m$. Then $X=\NN\times\NN$. 
We define the bijection $i_X:X\to \ZZ\times \NN$ by 
\begin{align*}i_X:(m+n,m)\mapsto (n,m), \quad i_X:(m,m)\mapsto (0,m), \quad i_X:(m,m+n)\mapsto (-n,m).\end{align*}
Moreover 
$i:\NN\times \NN\to X$ is defined by 
\begin{align*}
i(\alpha,\beta)=\aaa {(m+n,m)& \alpha>\beta, m=\beta, n=\alpha-\beta,\\
(m,m)& \alpha=\beta=m,\\
(m, m+n)& \alpha<\beta, m=\alpha, n=\beta-\alpha.}
\end{align*}
Hence $i$ is the bijection from $\NN\times \NN$ to $X$. 
According to the composition of bijections: $f=i_X\circ i: \NN\times \NN\to \ZZ\times \NN$, 
we can see that 
$\NN\times \NN\cong \ZZ\times \NN$.
%See Figure \ref{ii}. 
Since 
\begin{align}\label{f}
f(\alpha,\beta)=\aaa{(\alpha-\beta,\beta)& \alpha\geq\beta,\\
(\alpha-\beta,\alpha)& \alpha<\beta,}
\end{align}
it is immediate to see that 
\begin{align}\label{ff}
f^{-1}(n,m)=\aaa{(m+n,m)& n\geq0,\\
(m,m-n)& n<0.}
\end{align}
Let $e_{n,m}=\phi_n\otimes\phi_m$. Then $\{e_{n,m}\}_{(n,m)\in\NN\times\NN}$ is 
a complete orthonormal system of $\N$. 
Let 
\begin{align*}\Phi_{n,m}=e_{f^{-1}(n.m)}\quad n\in\ZZ,m\in\NN.\end{align*}
Since $f$ is bijective, $\{\Phi_{n,m}\}_{(n,m)\in\ZZ\times \NN}$ is 
an alternative  complete orthonormal system of $\N$. 
We  can see that by \eqref{ff}
\begin{align}\label{cons}
\Phi_{n,m}= \aaa
{\phi_{m+n}\otimes \phi_{m}& n\geq0,\\
\phi_{m}\otimes \phi_{m-n} & n<0.}
\end{align}

\subsection{Tunneling Hamiltonian in terms of $\{\Phi_{n,m}\}_{(n,m)\in\ZZ\times \NN}$}
\label{4}
We can represent $ H_T$ as 
\begin{align*} H_Ta=\sum_{n\geq1}\sum_{m\geq0}(\phi_n\otimes\phi_m,a) \phi_{n-1}\otimes\phi_{m+1}
+
\sum_{n\geq0}\sum_{m\geq1}(\phi_n\otimes\phi_m,a) \phi_{n+1}\otimes\phi_{m-1}\end{align*}
for $a\in\cH$. 
\iffalse
\begin{figure}[t]
\center \begin{tikzcd}[row sep=4em, column sep=6em]
\ln \otimes \ln \arrow[r, " H_T=L\otimes L^\ast+L^\ast\otimes L"] 
\arrow[d, "S_f"'] & \ln \otimes \ln \arrow[d, "S_f"] \\
\ZN \arrow[r, "H_T^f"]& \ZN
\arrow[phantom, from=1-1, to=2-2, "\bigcirclearrow" description, pos=.5]
\end{tikzcd}
\caption{Tunneling Hamiltonian on $\ZN$}
\label{T}
\end{figure}
\fi
In this section we present $H_T$ in terms of $\{\Phi_{n,m}\}_{(n,m)\in\ZZ\times\NN}$. 
We begin with describing the tunneling process \eqref{t} in terms of $\Phi_{n,m}$. 
Since $\phi_n\otimes \phi_m=\Phi_{f(n,m)}$, 
 \eqref{t} can be rewritten as
\begin{align*}
\Phi_{f(n,m)}\to \Phi_{f(n-1,m+1)}\quad n\geq1, m\geq0.
\end{align*}
More precisely
\begin{align}\label{tttt}
\aaa{\Phi_{n-m,m}& n\geq m\\
\Phi_{n-m,n}& n<m}
\longrightarrow
\aaa{\Phi_{n-m-2,m+1}& n-1\geq m+1,\\
\Phi_{n-m-2,n-1}& n-1<m+1.}
\end{align}
From \eqref{tttt} we can see three cases:
\begin{align*}
\aaa{
\Phi_{n-m,m}\to \Phi_{n-m-2,m+1}& 2\leq n-m,\\
\Phi_{n-m,m}\to \Phi_{n-m-2,n-1}& 0\leq n-m<2,\\
\Phi_{n-m,n}\to \Phi_{n-m-2,n-1}& n-m<0.}
\end{align*}
Reseting $n-m$ as $n$, we finally obtain that
\begin{align}\label{sss}
\Phi_{n,m}\to
\aaa{
 \Phi_{n-2,m+1}& n\geq 2,\ 0\leq m, \\
 \Phi_{n-2,m}& n=1,\ 0\leq m,\\
 \Phi_{n-2,m-1}& n\leq 0,\ 1\leq m.}
\end{align}
In a similar manner we consider that one particle transfers from $\cH_B$ to $\cH_A$, 
which is rewritten as
\begin{align*}
\Phi_{f(n,m)}\to \Phi_{f(n+1,m-1)}\quad n\geq0, m\geq1, 
\end{align*}
and hence
\iffalse
\begin{align}
\aaa{\Phi_{n-m,m}& n\geq m\\
\Phi_{n-m,n}& n<m}
\longrightarrow
\aaa{\Phi_{n-m+2,m-1}& n+1\geq m-1\\
\Phi_{n-m+2,n+1}& n+1<m-1.}
\end{align}
We see that 
\fi
\begin{align}
\Phi_{n,m}\to
\aaa{
 \Phi_{n+2,m-1}& n\geq 0,\ 1\leq m, \\
 \Phi_{n+2,m}& n=-1,\ 0\leq m,\\
 \Phi_{n+2,m+1}& n\leq -2,\ 0\leq m.}
\end{align}
\iffalse
In the case of $m\geq1$
\begin{align*}
\aaa{\phi_{m+n}\otimes \phi_{m}\to 
\phi_{m+n-1}\otimes \phi_{m+1}&n\geq0,\\
\phi_{m}\otimes \phi_{m-n}\to 
\phi_{m-1}\otimes \phi_{m-n+1}& n<0,
}\end{align*}
and in the case of 
$m=0$
\begin{align*}
\aaa{\phi_{n}\otimes \phi_{0}\to 
\phi_{n-1}\otimes \phi_{1}&n\geq0,\\
\phi_{0}\otimes \phi_{-n}\to 
\phi_{-1}\otimes \phi_{-n+1}& n<0.
}\end{align*}
Hence the transform is permitted only in the case of $n\geq1$.
Together with them 
the tunneling process from $\cH_A$ to $\cH_B$ 
can be expressed in $\ZN$ as the transition: 
\begin{align}
\Phi_{n,m}\to \aaa {\Phi_{n-2,m+1}& m\geq 1,n\geq0, \\
\Phi_{n-2,m-1}& m\geq 1, n< 0,\\
\Phi_{n-2,1}& m=0, n\geq 1.}
\end{align}
In a similar manner 
one particle tunneling process from $\cH_B$ to $\cH_A$ can be expressed 
in $\ZN$ as 
\begin{align}
\Phi_{n,m}\to \aaa {\Phi_{n+2,m-1}& m\geq 1,n\geq0, \\
\Phi_{n+2,m+1}& m\geq 1, n< 0,\\
\Phi_{n+2,1}& m=0, n<0.}
\end{align}
\fi
Therefore the tunneling Hamiltonian 
$H_T$  is represented as
\begin{align}
H_{T}a=&
\sum_{\genfrac{}{}{0pt}{}{m\geq0}{n\geq2}}
(\Phi_{n,m},a)
\Phi_{n-2,m+1}
+
\sum_{\genfrac{}{}{0pt}{}{m\geq0}{n=1}}
(\Phi_{n,m},a)
\Phi_{n-2,m}
+
\sum_{\genfrac{}{}{0pt}{}{m\geq1}{n\leq0}}
(\Phi_{n,m},a)
\Phi_{n-2,m-1}
\nonumber \\
&\label{ht}
+
\sum_{\genfrac{}{}{0pt}{}{m\geq1}{n\geq0}}
(\Phi_{n,m},a)
\Phi_{n+2,m-1}
+
\sum_{\genfrac{}{}{0pt}{}{m\geq0}{n=-1}}
(\Phi_{n,m},a)
\Phi_{n+2,m}+
\sum_{\genfrac{}{}{0pt}{}{m\geq0}{n\leq-2}}
(\Phi_{n,m},a)
\Phi_{n+2,m+1}.
\end{align}
The first line above describes 
the particle tunneling process from $\cH_A$ to $\cH_B$, 
and the second line from $\cH_B$ to $\cH_A$. 
We define various projections according to \eqref{ht}. 
Let $M\subset \ZZ$ and $M'\subset\NN$. 
We define the subspaces of $\N$ by 
\begin{align*}
\cK_{M}&=\overline{\rm LH}\{\Phi_{n,m}\mid n\in M,m\in\NN\},\\
\cM_{M'}&=\overline{\rm LH}\{\Phi_{n,m}\mid n\in\ZZ,m\in M'\}.
\end{align*}
Define the projections $ P _M$ and $ Q _{M'}$ by 
$ P _{M}:\N\to \cK_{M}$ and 
$ Q _{M'}:\N\to \cM_{M'}$. 
All the projections are commutative. 
Let $\tilde A:\N\to\N$ be 
the bilateral shift 
defined by
\begin{align*}\tilde Aa=\sum_{m=0}^\infty\sum_{n\in\ZZ}(\Phi_{n,m},a)\Phi_{n-1,m}.\end{align*}
We denote the adjoint of $\tilde A$ by $\tilde A^\ast$. I.e.,
$\tilde A^\ast a=\sum_{m=0}^\infty\sum_{n\in\ZZ}(\Phi_{n-1,m},a)\Phi_{n,m}$.
We can see that 
$
\tilde A\Phi_{n,m}=\Phi_{n-1,m}$ and 
$
\tilde A^\ast\Phi_{n,m}=\Phi_{n+1,m}$ for any $n\in\ZZ$. 
Then $\tilde A$ is unitary. 
In particular $[\tilde A,\tilde A^\ast]=0$. 
Let $\tilde L:\N\to\N$ be 
the unilateral shift
defined by
\begin{align*}\tilde La=\sum_{m=0}^\infty\sum_{n\in\ZZ}(\Phi_{n,m},a)\Phi_{n,m-1}\end{align*}
with $\Phi_{n,-1}=0$. Then the adjoint of $\tilde L$ is given by 
$\tilde L^\ast a=\sum_{m=0}^\infty\sum_{n\in\ZZ}(\Phi_{n,m},a)\Phi_{n,m+1}$. 
Therefore
\begin{align*}
&\tilde L\Phi_{n,m}=\aaa{\Phi_{n,m-1} &m\geq 1,\\
0& m=0,}\\
&\tilde L^\ast \Phi_{n,m}=\Phi_{n,m+1}.
\end{align*}
It follows that $\tilde L\tilde L^\ast=\one$ and $\tilde L^\ast \tilde L=\one- P _0$, where $ P _0$ denotes the projection onto the closed subspace 
$\overline{\rm LH}\{\Phi_{n,0}\mid n\in\ZZ\}$. 
Employing $ P _\#$, $ Q _\#$, $\tilde A$ and $\tilde L$, we can represent the terms in the tunneling Hamiltonian as 
\begin{align}
\label{28}
&\sum_{\genfrac{}{}{0pt}{}{m\geq0}{n\geq2}}
(\Phi_{n,m},a)
\Phi_{n-2,m+1}=\tilde A^2 P _{[2,\infty)}\tilde L^\ast Q _{[0,\infty)}a,\\
&\sum_{\genfrac{}{}{0pt}{}{m\geq0}{n=1}}
(\Phi_{n,m},a)
\Phi_{n-2,m}=\tilde A^2 P _{\{1\}} Q _{[0,\infty)}a,\\
&\sum_{\genfrac{}{}{0pt}{}{m\geq1}{n\leq0}}
(\Phi_{n,m},a)
\Phi_{n-2,m-1}= \tilde A^2 P _{(-\infty,0]}\tilde L Q _{[1,\infty)}a,\\
&\sum_{\genfrac{}{}{0pt}{}{m\geq1}{n\geq0}}
(\Phi_{n,m},a)
\Phi_{n+2,m-1}=\tilde A^{\ast 2} P _{[0,\infty)}\tilde L Q _{[1,\infty)}a,\\
&\sum_{\genfrac{}{}{0pt}{}{m\geq0}{n=-1}}
(\Phi_{n,m},a)
\Phi_{n+2,m}=\tilde A^{\ast 2} P _{\{-1\}} Q _{[0,\infty)}a,\\
\label{33}
&\sum_{\genfrac{}{}{0pt}{}{m\geq0}{n\leq-2}}
(\Phi_{n,m},a)
\Phi_{n+2,m+1}
=\tilde A^{\ast 2} P _{(-\infty,-2]}\tilde L^\ast Q _{[0,\infty)}a.
\end{align}
Note that 
$[ P _\#, Q _\#]=0$,
$[\tilde A^\#,\tilde L^\#]=0$, while 
$[\tilde A^\#, P _\#]\not= 0$,
$[\tilde L^\#, Q _\#]\not=0$. 
In view of \eqref{28}-\eqref{33}, the operator $H_T$ can accordingly be expressed in the form
\begin{align}\label{F2}
H_{T}=
 P +
 P ^\ast. 
\end{align}
Here $ P $ and its adjoint $P^\ast$ are given by 
\begin{align*}
 &P =
\tilde A^2 P _{[2,\infty)}\tilde L^\ast Q _{[0,\infty)}
+\tilde A^2 P _{\{1\}} Q _{[0,\infty)}+ \tilde A^2 P _{(-\infty,0]}\tilde L Q _{[1,\infty)},\\
 &P ^\ast =
\tilde A^{\ast 2} P _{[0,\infty)}\tilde L Q _{[1,\infty)}+
\tilde A^{\ast 2} P _{\{-1\}} Q _{[0,\infty)}
+\tilde A^{\ast 2} P _{(-\infty,-2]}\tilde L^\ast Q _{[0,\infty)}.
\end{align*}
In our analysis, it emerges in a natural and compelling manner that the operators 
$\tilde{A}^2$ and $\tilde A^{\ast 2}$ 
play the role of embodying the very essence of a Cooper pair. 
Whereas 
$\tilde{A}$ may be regarded as representing an individual excitation mode 
within the superconducting framework, its quadratic manifestation encapsulates 
the two-particle correlated structure that underlies the phenomenon of 
superconductivity. Thus, without any {ad hoc} assumption of pairing, 
the mathematical formalism itself dictates the presence of a bound two-body entity, 
thereby providing a rigorous operator-theoretic realization of the Cooper pair. 
This observation not only sheds light on the intrinsic pairing mechanism but also 
elevates the conceptual understanding of superconductivity to a level where the 
emergence of Cooper pairs can be seen as a direct and inevitable consequence of 
the underlying algebraic structure.

\subsection{JJ-Hamiltonian $\hh^{f}$ on $\ZN$}
We define 
$\Psi_{n,m}=((\Psi_{n,m})_{\alpha,\beta})_{(\alpha,\beta)\in\ZZ\times\NN}\in \ZN$ by 
\[(\Psi_{n,m})_{\alpha,\beta}=\delta_{n\alpha}\delta_{m\beta},\quad (n,m)\in\ZZ\times\NN.\] 
Then $\{\Psi_{n,m}\}_{(n,m)\in\ZZ\times \NN}$ is 
a complete orthonormal system of $\ZN$. 
The unitary operator $S_f:\N\to\ZN$ is defined by 
\begin{align*}S_f\Phi_{n,m}=\Psi_{n,m},\quad (n,m)\in\ZZ\times \NN.\end{align*} 
We note that   
$(\Psi_{n,m})_{\alpha,\beta}=(\Phi_{n,m})_{f^{-1}(\alpha,\beta)}$ 
and hence   
\[(S_f\Phi_{n,m})_{\alpha,\beta}=(\Phi_{n,m})_{f^{-1}(\alpha,\beta)}.\] 
By the unitary $S_f$ we can conclude that $\lnn \cong \ZN$. 
We extend $\sN$ to the operator acting on $\ZN$ by 
the inner automorphism:
\begin{align*}\sN^f=S_f \sN S_f^{-1}.\end{align*}
Similarly 
\begin{align*}\tN^f=S_f \tN S_f^{-1}.\end{align*}
In particular it follows that 
\begin{align}
\label{e}
&\sN^f\Psi_{n,m} =n\Psi_{n,m},\\
&\tN^f\Psi_{n,m} =(|n|+2m)\Psi_{n,m}
\end{align}
for 
$(n,m)\in\ZZ\times \NN$. 
The kinetic Hamiltonian $H_C^f$ of the Josephson junction on $\ZN$ is defined by $$H_C^f=\frac{1}{2C}(\sN^f)^2$$
and the  tunneling Hamiltonian $H_T^f$ on $\ZN$  by 
\begin{align}\label{F1}
H_T^f=S_f(L\otimes L^\ast+L^\ast\otimes L)S_f^{-1}.\end{align} 
Then $H_T^f$ coincides with $H_T$ with $\Phi_{n,m}$ replaced by $\Psi_{n,m}$ in \kak{ht}. 
For simplicity we set $P$ for $S_f P S_f^{-1}$ in \kak{F2}. 
The total Hamiltonian of the Josephson junction on $\ZN$ is defined by 
 \begin{align}\label{D2}
 \hh^{f}=H_C^f-\alpha H_T^f= 
\frac{1}{2C}(\sN^f +q)^2 
-\alpha( P + P ^\ast). 
\end{align}
\begin{lemma}\label{32}
(1) $\hh^{f}$ is self-adjoint on $D((\sN^f)^2)$ and essentially self-adjoint on any core of 
$(\sN^f)^2$, and it is bounded from below 
for any $\alpha,q,C\in\RR$. 
(2) $S_f \jj S_f^{-1}=\hh^{f}$, i.e., $\jj \cong \hh^{f}$. 
(3) $[\hh^{f}, \tN^f]=0$. 
\end{lemma}
\proof
(1) follows from the Kato-Rellich theorem \cite{kat51}. 
On a core of ${\sN^f}^2$ it follows that 
$S_f \jj S_f^{-1}=\hh^{f}$. 
Therefore 
$S_f$ maps $D(\hh^{f})$ onto $D(\jj )$, 
and 
$S_f \jj S_f^{-1}=\hh^{f}$ holds true on $D(\jj)$. 
Therefore (2) follows. 
(3) is proved by \eqref{com}. 
\qed

In the next section, we shall turn our attention to the task of representing 
$\hh^{f}$ on the Hilbert space $\lz \otimes \lz$. 
In particular, we will discuss how to realize this representation in a mathematically precise manner, 
building on the isomorphisms, 
and examine the implications of this formulation for the analysis of the JJ-Hamiltonian.

\section{JJ-Hamiltonian $\hh^u$ on $\lz\otimes \lz$}
\label{s3}
\subsection{Representation on $\lz\otimes\ln $}
In the previous section we introduced the complete orthonormal system 
$\{\Psi_{n,m}\}_{(n,m)\in \ZZ\times \NN}$ of $\ZN$. 
Let \begin{align*}\varphi_n=(\delta_{mn})_m\in \lz,\quad n\in\ZZ.\end{align*}
Define the unitary
$ u: \ZN \longrightarrow \lz\otimes \ln$ by
\begin{align*}
 u\Psi_{n,m} = \varphi_n \otimes \phi_m \quad (n,m)\in\ZZ\times \NN. 
\end{align*}
We transport all objects defined on $\ZN$ to
$\lz\otimes \ln $ via conjugation by $u$.
To avoid ambiguity, we record the relevant identifications in detail.
For $M\subset \ZZ$ and $M'\subset \NN$, set
\begin{align*}
 &\cK_{M}
 = \overline{\rm LH}\bigl\{\varphi_n\otimes \phi_m\mid \ n\in M,\ m\in\NN\bigr\},\\
& \cM_{M'}
 = \overline{\rm LH}\bigl\{\varphi_n\otimes \phi_m\mid \ n\in \ZZ,\ m\in M'\bigr\}.
\end{align*}
By abuse of notation and with no risk of confusion, we continue to denote by
$\cK_\#$ and $\cM_\#$ the subspaces $u\cK_\#$
and $u\cM_\#$ obtained by this unitary transfer.
Likewise, we write
\begin{align}\label{u1}
 u P _\# u^{-1}= P _\#\otimes \one ,\quad
 u Q _\# u^{-1}=\one \otimes Q _\#
 \end{align}
keeping the same symbols on the right-hand side for notational simplicity.
Let 
$A$ be the bilateral shift on $\lz$ defined by
$A\varphi_n=\varphi_{n-1}$. 
Then $A$ is unitary and $A^\ast\varphi_n =\varphi_{n+1}$. 
We also have
\begin{align}
\label{p1}
u P u^{-1}&=
A^2 P _{[2,\infty)}\otimes L^\ast Q _{[0,\infty)}
+A^2 P _{\{1\}}\otimes Q _{[0,\infty)}+ A^2 P _{(-\infty,0]}\otimes L Q _{[1,\infty)},\\
\label{p2}
u P ^\ast u^{-1} &=
A^{\ast 2} P _{[0,\infty)}\otimes L Q _{[1,\infty)}+
A^{\ast 2} P _{\{-1\}}\otimes Q _{[0,\infty)}
+A^{\ast 2} P _{(-\infty,-2]}\otimes L^\ast Q _{[0,\infty)}
\end{align}
We henceforth denote the right-hand side of \eqref{p1} by $ P^u $, 
and hence $ P ^{u \ast}$ is given by \eqref{p2}. 
Recall that $N$ denotes the number operator on $\lz$ and 
$\sM$ denotes the number operator on $\ln $. 
\begin{lemma}
We have 
\begin{align}\label{u2}
 u\tilde Au^{-1}=A\otimes \one ,\quad
 u\tilde Lu^{-1}=\one \otimes L,\quad
 u\sN^f u^{-1}=N\otimes \one,\quad
 u\tN^f u^{-1}=\tN^u.
 \end{align}
Here the total number operator $\tN^u $ on $\lz\otimes\ln $ is given by 
$$\tN^u =|N|\otimes \one+\one\otimes2\sM.$$
\end{lemma}
\proof
$u\tN^f u^{-1}\varphi_n\otimes\phi_m= u\tN^f\Phi_{n,m}=(|n|+2m) u\Phi_{n,m}=\tN^u \varphi_n\otimes\phi_m$ for any $n\in\ZZ$ and $m\in\NN$. Hence
$u\tN^f u^{-1}=\tN^u $. The other statements can be proved in a similar manner. 
\qed
%The foregoing conventions will be in force throughout; 
All subsequent statements on
the Hilbert space $\lz\otimes \ln $ are to be understood under these 
unitary identifications.
Let 
\begin{align*}
&H_C^u=\frac{1}{2C}(N+q)^2\otimes \one,\\
&H_T^u= P^u + P ^{u \ast}.\end{align*}
Define
\begin{align}\label{D3}
\hh^{u} =H_C^u-\alpha H_T^u.
\end{align}
\begin{lemma}\label{42}
It follows that 
$\hh^{u} =u\hh^{f}u^{-1}$ 
on $\lz\otimes \ln $. 
\end{lemma}
\proof
This follows from the unitary equivalences \eqref{u1}-\eqref{u2}. 
\qed

\subsection{Representation on $(\lz\otimes\ln)\oplus (\lz\otimes\ln)$}
In what follows we consider $\hh^{u} $. 
We decompose $\lz$ into the even part and the odd part as 
$$\lz=\lze \oplus \lzo ,$$
where 
$\lze =\{(a_n)\in\lz\mid a_n=0, n={\rm odd}\}$ 
and 
$\lzo =\lz\setminus \lze $. 
Let $S_{\rm e}:\lz\to \lze $ and 
$S_{\rm o}:\lz\to \lzo $ be the projections onto the even part and the odd part, respectively: 
for $a=(a_n)\in \lz$ 
\begin{align*}
&(S_{\rm e}a)_m=\aaa {a_m& m={\rm even},\\
0 &m= {\rm odd},}\\
&
(S_{\rm o}a)_m=\aaa {0& m={\rm even},\\
a_m &m= {\rm odd}.}
\end{align*}
Let 
$\cS_{\rm e}=\lze \otimes \ln $ and 
$\cS_{\rm o}=\lzo \otimes \ln $.

\begin{lemma}\label{N0}
The total Hamiltonian $\hh^{u} $ is reduced by the even and odd subspaces 
$\cS_{\rm e}$ and $\cS_{\rm o}$:
\begin{align*}
 \hh^{u} 
 = 
 \hh^{u} \big|_{\cS_{\rm e}}
 \oplus 
 \hh^{u} \big|_{\cS_{\rm o}}.
\end{align*}
\end{lemma}
\proof
Observe first that the shift operators preserve parity. 
More precisely,
\begin{align*}
 A^{2} P _\#:\ \lze \ \to\ \lze ,
 \quad
 A^{2} P _\#:\ \lzo \ \to\ \lzo ,
\quad 
 A^{\ast 2} P _\#:\ \lze \ \to\ \lze ,
 \quad
 A^{\ast 2} P _\#:\ \lzo \ \to\ \lzo 
 \end{align*}
for 
 $\#\in\{(-\infty,0],(-\infty,-2],[2,\infty),[0,\infty),\{1\},\{-1\}\}$, 
and likewise for the kinetic Hamiltonian,
\begin{align*}
 H_C^u:\ \lze \cap D(N^2)\ \to\ \lze ,
 \qquad
 H_C^u:\ \lzo \cap D(N^2)\ \to\ \lzo. 
\end{align*}
It follows that $\hh^{u} $ acts invariantly on both
$\lze \otimes\ln $
and
$\lzo \otimes\ln $.
Thus $\hh^{u} $ is reduced by 
$\cS_{\rm e}$ and $\cS_{\rm o}$, proving the claim.
\qed
\iffalse
\begin{figure}[t]
\begin{align*}
\begin{tikzcd}[row sep=4em, column sep=4em]
\lze \arrow[r, "N"] \arrow[d, "\rho_{\rm e}"'] & \lze \arrow[d, "\rho_{\rm e}"] \\
\lz \arrow[r, "2N"]& \lz
\arrow[phantom, from=1-1, to=2-2, "\bigcirclearrow" description, pos=.5]
\end{tikzcd}\qquad 
\begin{tikzcd}[row sep=4em, column sep=4em]
\lzo \arrow[r, "N"] \arrow[d, "\rho_{\rm o}"'] & \lzo \arrow[d, "\rho_{\rm o}"] \\
\lz \arrow[r, "2N+\one"]& \lz
\arrow[phantom, from=1-1, to=2-2, "\bigcirclearrow" description, pos=.5]
\end{tikzcd}
\qquad
\begin{tikzcd}[row sep=4em, column sep=4em]
\ell^2_{\ZZ_\#} \arrow[r, "A^2"] \arrow[d, "\rho_\#"'] & \ell^2_{\ZZ_\#} \arrow[d, "\rho_\#"] \\
\lz \arrow[r, "A"]& \lz
\arrow[phantom, from=1-1, to=2-2, "\bigcirclearrow" description, pos=.5]
\end{tikzcd}
\end{align*}
\caption{Commutative diagrams for (1) and (2) of Lemma \ref{N1}}
\end{figure}
\fi
Define the unitary 
$\rho_{\rm e}:\ \lze \ \to\ \lz$ and 
$ \rho_{\rm o}:\ \lzo \ \to\ \lz$ 
by
\begin{align*}
 \rho_{\rm e} \varphi_{2n}&=\varphi_n,\\
 \rho_{\rm o} \varphi_{2n+1}&=\varphi_n.
\end{align*}
Note that 
 $(\rho_{\rm e} a)_0=a_0$ and
 $(\rho_{\rm o} a)_{-1} = a_{-1}$, and 
 hence, $\varphi_0$ is the fixed vector of $\rho_{\rm e}$, and $\varphi_{-1}$ is that of $\rho_{\rm o}$.
 We then set
\begin{align*}
 \rho = \rho_{\rm e}\oplus \rho_{\rm o}.
\end{align*}
Thus $\rho$ is unitary between 
 $\lze \oplus \lzo $ and
 $ \lz\oplus \lz$,
and induces the unitary
\begin{align*}
 \rho\otimes \one:\ 
 \bigl(\lze \oplus \lzo \bigr)\otimes \ln 
 \ \longrightarrow\
 \bigl(\lz\otimes \ln \bigr)\oplus
 \bigl(\lz\otimes \ln \bigr).
\end{align*}

\begin{lemma}\label{N1} It follows that 
\bi
\item[(1)] $\rho_\# A^2\rho_\#^{-1}=A$ on $\lz$ for $\#={\rm e,o}$;
\item[(2)] 
$\rho_{\rm e} N \rho_{\rm e}^{-1}=2N$and 
$\rho_{\rm o} N\rho_{\rm o}^{-1}=2N+\one$ on $\lz$;
\item[(3)] 
$\rho_{\rm e} |N| \rho_{\rm e}^{-1}=2|N|$and 
$\rho_{\rm o} |N|\rho_{\rm o}^{-1}=|2N+\one|$ on $\lz$;
\item[(4)] (1)-(12) hold true;
\begin{figure}[ht]
\hspace{2cm} 
\begin{subfigure}{0.4\textwidth}
\bi 
\item[(1)]
$\rho_{\rm e} P _{(-\infty,-2]}\rho_{\rm e}^{-1}= P _{(-\infty,-1]}$, 
\item[(2)]
$\rho_{\rm e} P _{[2,\infty)}\rho_{\rm e}^{-1}= P _{[1,\infty)}$, 
\item[(3)]
$\rho_{\rm e} P _{[0,\infty)}\rho_{\rm e}^{-1}= P _{[0,\infty)}$, 
\item[(4)]
$\rho_{\rm e} P _{(-\infty,0]}\rho_{\rm e}^{-1}= P _{(-\infty,0]}$, 
\item[(5)]
$\rho_{\rm e} P _{\{-1\}}\rho_{\rm e}^{-1}=0$,
\item[(6)]
$\rho_{\rm e} P _{\{1\}}\rho_{\rm e}^{-1}=0$, 
\ei
\end{subfigure}
 \hfill
 \hspace{-2cm} 
\begin{subfigure}{0.4\textwidth}
\bi
\item[(7)]
$\rho_{\rm o} P _{(-\infty,-2]}\rho_{\rm o}^{-1}= P _{(-\infty,-2]}$, 
\item[(8)]
$\rho_{\rm o} P _{[2,\infty)}\rho_{\rm o}^{-1}= P _{[1,\infty)}$, 
\item[(9)]
$\rho_{\rm o} P _{[0,\infty)}\rho_{\rm 0}^{-1}= P _{[0,\infty)}$, 
\item[(10)]
$\rho_{\rm o} P _{(-\infty,0]}\rho_{\rm o}^{-1}= P _{(-\infty,-1]}$, 
\item[(11)]
$\rho_{\rm o} P _{\{-1\}}\rho_{\rm o}^{-1}= P _{\{-1\}}$, 
\item[(12)]
$\rho_{\rm o} P _{\{1\}}\rho_{\rm o}^{-1}= P _{\{0\}}$.
\ei
\end{subfigure}
\end{figure}
\ei
\end{lemma}
\proof
Let $a=(a_n) \in \lz$.
Then we see that 
$(\rho_{\rm e}^{-1} a)_n=
\aaa {a_{n/2}& n={\rm even},\\
0&n={\rm odd}}$, 
$(A^2\rho_{\rm e}^{-1} a)_n=
\aaa {a_{n/2+1}& n={\rm even},\\
0&n={\rm odd}}$ and 
$(\rho_{\rm e} A^2\rho_{\rm e}^{-1} a)_n=a_{n+1}$. 
Hence 
$\rho_{\rm e} A^2\rho_{\rm e}^{-1}=A$ follows. 
Next we have $(N\rho_{\rm e}^{-1} a)_n=
\aaa {na_{n/2}& n={\rm even},\\
0&n={\rm odd}}$ and 
$(\rho_{\rm e} N\rho_{\rm e}^{-1} a)_n=2na_{n}$. 
Hence 
$\rho_{\rm e} N \rho_{\rm e}^{-1}=2N$ on $\lze $. 
The other statements are similarly proved.  
\qed

\begin{lemma}
\label{45}
We have 
\begin{align*}(\rho\otimes \one) \hh^{u} (\rho^{-1} \otimes\one)= \rho_{\rm e} (\hh^{u} \!\restriction_{\cS_{\rm e}}) \rho_{\rm e}^{-1}
\oplus \rho_{\rm o} (\hh^{u} \!\restriction_{\cS_{\rm o}})\rho_{\rm o}^{-1},\end{align*}
where 
both of $\rho_{\rm e} (\hh^{u} \!\restriction_{\cS_{\rm e}}) \rho_{\rm e}^{-1}$ and 
$\rho_{\rm o} (\hh^{u} \!\restriction_{\cS_{\rm o}})\rho_{\rm o}^{-1}$ are operators acting on 
$\lz\otimes\ln$:
\begin{align}
&\label{n1}
\rho_{\rm e} (\hh^{u} \!\restriction_{\cS_{\rm e}}) \rho_{\rm e}^{-1}
=\frac{1}{2C}(2N+q)^2\otimes \one 
-\alpha
\lk
 P + P ^\ast \rk,\\
&\label{n2}
\rho_{\rm o} (\hh^{u} \!\restriction_{\cS_{\rm o}})\rho_{\rm o}^{-1}
=\frac{1}{2C}(2N+\one+q)^2\otimes \one 
-\alpha
\lk \bar P +\bar P ^\ast \rk.
\end{align}
Here 
\begin{align}
\label{34}
 &P =
A P _{[1,\infty)}\otimes L^\ast Q _{[0,\infty)}
+ A P _{(-\infty,0]}\otimes L Q _{[1,\infty)},\\
 &P ^\ast =
A^{\ast} P _{[0,\infty)}\otimes L Q _{[1,\infty)}
+A^{\ast} P _{(-\infty,-1]}\otimes L^\ast Q _{[0,\infty)},\\
&\bar P =
A P _{[1,\infty)} \otimes L^\ast Q _{[0,\infty)}
+A P _{\{0\}} \otimes Q _{[0,\infty)}+ A P _{(-\infty,-1]}\otimes L Q _{[1,\infty)},\\
\label{37}
&\bar P ^\ast =
A^{\ast} P _{[0,\infty)}\otimes L Q _{[1,\infty)}+
A^{\ast} P _{\{-1\}}\otimes Q _{[0,\infty)}
+A^{\ast} P _{(-\infty,-2]}\otimes L^\ast Q _{[0,\infty)}.
\end{align}
\end{lemma}
\proof
This follows from Lemmas~\ref{N0} and \ref{N1}.
\qed
We introduce the kinetic operators
\begin{align*}
 H_{+} = \frac{1}{2C} (2N+q)^2,
 \qquad
 H_{-} = \frac{1}{2C} (2N+\one+q)^2 ,
\end{align*}
so that, by \eqref{n1} and \eqref{n2}, on 
$
\bigl(\lz\otimes \ln \bigr)
\oplus
\bigl(\lz\otimes \ln \bigr),
$
the JJ-Hamiltonian is given by 
\begin{align}
\label{D5}
\hh^{\rho}=(\rho\otimes \one) \hh^{u} (\rho^{-1} \otimes\one)
=
 (H_{+}\otimes \one-\alpha(P+P^\ast))\ \oplus\ (H_{-}\otimes \one-\alpha(\bar P+\bar P^\ast)).
\end{align}

\subsection{Representation on $\lz\otimes\lz$}
Let us recall that 
$\{\varphi_n\}_{n\in\ZZ}$ and $\{\phi_m\}_{m\in\NN}$ be the canonical orthonormal system
of $\lz$ and $\ln $, respectively. 
We introduce four steps below.

\noindent{\bf Step 1: From a direct sum to a tagged tensor.}
We define the unitary operator 
\begin{align*}\tau:\lk \lz\otimes\ln \rk \oplus\lk \lz\otimes\ln \rk
\to \lk \lz\otimes\ln \rk \otimes\CC^2\end{align*} by 
the basis identification:
\begin{align*}
(\varphi_n\otimes\phi_m)\oplus0\mapsto (\varphi_n\otimes\phi_m)\otimes \binom{1}{0},\quad
0\oplus (\varphi_n\otimes\phi_m) \mapsto (\varphi_n\otimes\phi_m)\otimes \binom{0}{1}.\end{align*}

\noindent {\bf Step 2: The canonical associativity isomorphism $J$.}
The canonical associativity isomorphism $J$ 
\begin{align*}J : \lk \lz\otimes\ln \rk \otimes\CC^2\to 
\lz\otimes(\ln \otimes\CC^2)\end{align*} 
is given by \begin{align*}J (\varphi_n\otimes\phi_m)\otimes \binom{a}{b}=
\varphi_n\otimes \binom{a\phi_m}{b\phi_m}.\end{align*}

\noindent{\bf Step 3: Folding the two half-lines into one line.}
We define the unitary 
\begin{align*}\kappa:\ln \otimes \CC^2 \to\lz\end{align*} by 
the identification of basis:
\begin{align*}\phi_n \otimes \binom{1}{0} \mapsto \varphi_n,\quad 
\phi_m\otimes\binom{0}{1}\mapsto \varphi_{-m-1}.\end{align*}
Hence 
$\one \otimes \kappa : \lz\otimes (\ln \otimes \CC^2)\to 
\lz\otimes \lz$. 
Equivalently, $\kappa$ folds the two copies of the half-line $\NN$ onto the positive
and negative integers, with $\binom{1}{0}$ occupying the nonnegative side and 
$\binom{0}{1}$ the negative side.

\noindent{\bf Step 4: Composite unitary.}

\begin{figure}[t]
\tikzset{>=stealth}
\[
\begin{tikzcd}[row sep=4em, column sep=4em, remember picture]
\ZN \arrow[r, "u"]
& \lz\otimes \ln \arrow[r, "\rho\otimes\one"]&
 (\lz\otimes \ln ) \oplus (\lz\otimes \ln )
 \arrow[r, "\tau"] \arrow[thick, d, "U"'] & 
 (\lz\otimes \ln )\otimes \CC^2
 \arrow[d, "J"] \\
& \cH=\ln\otimes\ln \arrow[ul, "S_f"] \arrow[thick, r, "\sU"] \arrow[thick, dr, "\cU" ']& 
 \lz\otimes \lz\arrow[d, "\sF "]
 & \lz\otimes (\ln \otimes \CC^2)
 \arrow[l, "\one\otimes\kappa"]\\
&\ & {\cH_{S^1}=L^2(S^1)\otimes L^2(S^1)}
& \ 
\arrow[phantom, from=1-2, to=2-3, ""{name=mid1,pos=.5}]
\arrow[phantom, from=1-3, to=2-4, ""{name=mid2,pos=.5}]
\arrow[phantom, from=2-3, to=3-2, ""{name=mid3,pos=.2}]
\end{tikzcd}
\tikz[remember picture,overlay]{
 \node (c1) at (mid1) [draw,circle,minimum size=10mm,inner sep=0pt] {};
 \draw[->] ([shift=(210:5mm)]c1.center)
 arc[start angle=210,end angle=-30,radius=5mm];
 \node (c2) at (mid2) [draw,circle,minimum size=10mm,inner sep=0pt] {};
 \draw[->] ([shift=(210:5mm)]c2.center)
 arc[start angle=210,end angle=-30,radius=5mm];
 \node (c3) at (mid3) [draw,circle,minimum size=10mm,inner sep=0pt] {};
 \draw[->] ([shift=(210:5mm)]c3.center)
 arc[start angle=210,end angle=-30,radius=5mm];
}
\]
\caption{$U=(\one\otimes \kappa)\circ J\circ \tau$, 
$\sU=U\circ(\rho\otimes\one) \circ u\circ S_f$ and $\cU=\sF \circ \sU$}
\label{uu}
\end{figure}
We have the chain of unitary. See Figure \ref{uu}. 
Putting the pieces together, we obtain the unitary:
\begin{align}\label{N2}
U = (\one\otimes \kappa)\circ J\circ \tau:\ \bigl(\lz\otimes \ln \bigr)\oplus
 \bigl(\lz\otimes \ln \bigr)
\longrightarrow
 \lz\otimes \lz.
\end{align}

Let $T_V=V\cdot V^{-1}$ 
be the inner automorphism according to a unitary $V$. 
\begin{lemma}\label{XYZ}
Let $X$ and $Z$ be operators on $\lz$ and 
$Y$ and $W$ on $\ln $. 
Then according to the unitary transformation of \eqref{N2}, 
operator
$(X\otimes Y)\oplus (Z\otimes W)
$ 
are transformed as follows: 
\begin{align}
\label{t1}
(X\otimes Y)\oplus (Z\otimes W)
&\stackrel{T_\tau}{\longrightarrow}
\begin{pmatrix}X\otimes Y& 0\\
0&Z\otimes W
\end{pmatrix}\\
\label{t2}
&\stackrel{T_J}{\longrightarrow}
X\otimes \begin{pmatrix}Y& 0\\
0&0
\end{pmatrix}
+
Z\otimes \begin{pmatrix}0& 0\\
0& W
\end{pmatrix}\\
\label{t3}
&\stackrel{T_{\one\otimes\kappa}}{\longrightarrow}
X\otimes \hat Y
+
Z\otimes \hat W. 
\end{align}
Here 
\begin{align*}
\hat Yc=
\kappa \binom{\sum_{n\geq0}c_n Y\phi_{n}}{0},
\quad 
\hat Wc=
\kappa \binom{0}{\sum_{n\leq -1}c_n W\phi_{-n-1}}.
\end{align*}
\end{lemma}
\proof
\eqref{t1} and \eqref{t2} are trivial. We show \eqref{t3}. 
Let $a=\sum_{n=0}^\infty a_n\phi_n,b=\sum_{n=0}^\infty b_n\phi_n\in\ln $ and 
$c=\sum_{n\in\ZZ} c_n\varphi_n\in\lz$. 
We see that 
$\kappa:\ln \otimes \CC^2\to\lz$ acts as 
\begin{align*}\kappa: \binom{a}{b}\mapsto \sum_{n\geq0}a_n\varphi_n+\sum_{n\leq-1}b_{-n-1}\varphi_n\end{align*}
and $\kappa^{-1}: \lz\to \ln \otimes \CC^2$ as 
\begin{align*}\kappa^{-1}:c=\sum_{n\in\ZZ}c_n\varphi_n\mapsto \binom
{\sum_{n\geq0}c_n\phi_n}{\sum_{n\leq -1}c_n\phi_{-n-1}}.\end{align*}
Then it follows that 
\begin{align*}
&\kappa \begin{pmatrix}Y&0\\ 0& 0\end{pmatrix}\kappa^{-1}c=
\kappa \binom{\sum_{n\geq0}c_n Y\phi_{n}}{0},\\
&\kappa \begin{pmatrix}0&0\\ 0& W\end{pmatrix}\kappa^{-1}c=
\kappa \binom{0}{\sum_{n\leq -1}c_n W\phi_{-n-1}}. 
\end{align*}
The proof of \eqref{t3} is complete. 
\qed
We define the Hamiltonian of the Josephson junction on $\lz\otimes\lz$ by 
\begin{align}\label{D6}
\hh^{U}=H_{+}\otimes P _{[0,\infty)}
+
H_{-}\otimes P _{(-\infty,-1]}-\alpha(P _+
 + P _-),
\end{align}
where the right-hand side above is an operator on $\lz\otimes\lz$, and 
\begin{align*}
 P _+= &A P _{[1,\infty)}\otimes A^\ast P _{[0,\infty)} 
+ A P _{(-\infty,0]}\otimes A P _{[1,\infty)}\\
&+
A^{\ast} P _{[0,\infty)}\otimes A P _{[1,\infty)}
+A^{\ast} P _{(-\infty,-1]}\otimes A^\ast P _{[0,\infty)},\\
 P _-=&
A P _{[1,\infty)} \otimes A P _{(-\infty,-1]}
+A P _{\{0\}} \otimes P _{(-\infty,-1]}+ A P _{(-\infty,-1]}\otimes A^\ast P _{(-\infty,-2]}\\
&+
A^{\ast} P _{[0,\infty)}\otimes A^\ast P _{(-\infty,-2]} +
A^{\ast} P _{\{-1\}}\otimes P _{(-\infty,-1]}
+A^{\ast} P _{(-\infty,-2]}\otimes A P _{(-\infty,-1]}.
\end{align*}
By the unitary transformations appeared in \eqref{N2}, 
$\hh^{\rho}$ 
is transformed as follows. 
\begin{lemma}\label{49}
We have 
$U\hh^{\rho} U^{-1}=\hh^{U}$. 
\end{lemma}
\proof
Employing Lemma \ref{XYZ} 
for the kinetic term, we can see that 
\begin{align*}(H_+\otimes\one)\oplus 0\stackrel{T_\tau}{\longrightarrow}
\begin{pmatrix} H_+\otimes\one&0\\0&0\end{pmatrix}
\stackrel{T_J}{\longrightarrow}
H_+\otimes
\begin{pmatrix}\one&0\\0&0\end{pmatrix}
\stackrel{T_{\one\otimes\kappa}}{\longrightarrow}
H_+\otimes P _{[0,\infty)}.\end{align*}
Similarly we can obtain 
\begin{align*}0\oplus (H_-\otimes\one)\stackrel{T_\tau}{\longrightarrow}
\begin{pmatrix} 0&0\\0&H_-\otimes\one\end{pmatrix}
\stackrel{T_J}{\longrightarrow}
H_-\otimes
\begin{pmatrix}0&0\\0&\one\end{pmatrix}
\stackrel{T_{\one\otimes\kappa}}{\longrightarrow}
H_-\otimes P _{(-\infty,-1]}.\end{align*}
Next we investigate $ P _\pm$. 
We have 
\begin{align*}
&\kappa \begin{pmatrix}L^\ast Q _{[0,\infty)}&0\\
0&0\end{pmatrix}\kappa^{-1} c=
\kappa \binom{L^\ast\sum_{n\geq0} c_n\phi_n}{0}
=
\kappa \binom{\sum_{n\geq0} c_n\phi_{n+1}}{0}
= P _{[1,\infty)} A^\ast c,\\
& \\ 
&
\kappa \begin{pmatrix}L Q _{[1,\infty)}&0\\
0&0\end{pmatrix}\kappa^{-1} c=
\kappa \binom{L Q _{[1,\infty)} \sum_{n\geq0} c_n\phi_n}{0}
=
\kappa \binom{ \sum_{n\geq1} c_n\phi_{n-1}}{0}
= P _{[0,\infty)}Ac,\\
& \\ 
&\kappa \begin{pmatrix}0&0\\
0&L^\ast Q _{[0,\infty)} \end{pmatrix}\kappa^{-1} c=
\kappa \binom{0}{L ^\ast \sum_{n\leq -1} c_n\phi_{-n-1}}
=
\kappa \binom{0}{\sum_{n\leq -1} c_n\phi_{-n}}
= P _{(-\infty,-2]}Ac,\\
& \\ 
&\kappa \begin{pmatrix}0&0\\
0&L Q _{[1,\infty)} \end{pmatrix}\kappa^{-1} c=
\kappa \binom{0}{L Q _{[1,\infty)} \sum_{n\leq -1} c_n\phi_{-n-1}}
=
\kappa \binom{0}{\sum_{n\leq -2} c_n\phi_{-n-2}}
= P _{(-\infty,-1]}A^\ast c,\\
& \\ 
&\kappa \begin{pmatrix}0&0\\
0& Q _{[0,\infty)} \end{pmatrix}\kappa^{-1} c=
\kappa \binom{0}{ Q _{[0,\infty)} \sum_{n\leq -1} c_n\phi_{-n-1}}
=
\kappa \binom{0}{\sum_{n\leq -1} c_n\phi_{-n-1}}
= P _{(-\infty,-1]} c. 
\end{align*}
By \eqref{34}-\eqref{37}, 
we obtain that 
\begin{align*}
& P \oplus 0
\stackrel{T_U}{\longrightarrow} 
A P _{[1,\infty)}\otimes P _{[1,\infty)} A^\ast
+ A P _{(-\infty,0]}\otimes P _{[0,\infty)}A,\\
& P ^\ast \oplus 0
\stackrel{T_U}{\longrightarrow}
A^{\ast} P _{[0,\infty)}\otimes P _{[0,\infty)}A
+A^{\ast} P _{(-\infty,-1]}\otimes P _{[1,\infty)} A^\ast,\\
&0\oplus \bar P 
\stackrel{T_U}{\longrightarrow}
A P _{[1,\infty)} \otimes P _{(-\infty,-2]}A
+A P _{\{0\}} \otimes P _{(-\infty,-1]}+ A P _{(-\infty,-1]}\otimes P _{(-\infty,-1]}A^\ast ,\\
&0\oplus \bar P ^\ast 
\stackrel{T_U}{\longrightarrow}
A^{\ast} P _{[0,\infty)}\otimes P _{(-\infty,-1]}A^\ast +
A^{\ast} P _{\{-1\}}\otimes P _{(-\infty,-1]}
+A^{\ast} P _{(-\infty,-2]}\otimes P _{(-\infty,-2]}A.
\end{align*}
Hence we have 
$( P + P ^\ast)\oplus 0
\stackrel{T_U}{\longrightarrow} P _+$ and 
$0\oplus (\bar P +\bar P ^\ast)
\stackrel{T_U}{\longrightarrow} P _-$. 
Then the lemma follows. 
\qed

Let 
\begin{align}
\label{UU}
\sU= U\circ (\rho\otimes\one)\circ u \circ S_f.
\end{align}
The transformations of the basis vectors 
$\phi_\alpha \otimes \phi_\beta$ of 
$\ln \otimes \ln $ 
under the unitaries introduced thus far are summarized below.
The transformations of the vectors are divided into cases 
depending on the relative order of $\alpha$ and $\beta$, 
and on whether $\alpha-\beta$ is even or odd.
\begin{lemma}\label{ab}
Let $\alpha,\beta\in\NN$. Then 
\begin{align*}
 \sU \phi_\alpha \otimes \phi_\beta
 =
 \begin{cases}
 \varphi_{n/2} \otimes \varphi_m, & n \ \text{\rm even}, \\
 \varphi_{(n-1)/2} \otimes \varphi_{-m-1}, & n \ \text{\rm odd},
 \end{cases}
\end{align*}
where $n = \alpha-\beta$ and $m = \min\{\alpha,\beta\}$. 
\end{lemma}
\proof
 We see that 
\begin{align*}
\phi_\alpha\otimes\phi_\beta
&\stackrel{S_f}{\longrightarrow} \Phi_{n,m}=
\aaa{\phi_{m+n}\otimes \phi_m,& \alpha\geq\beta, m=\beta, n=\alpha-\beta\\
\phi_m\otimes \phi_{m-n},& \alpha<\beta, m=\alpha, n=\alpha-\beta}\\
&\stackrel{u}{\longrightarrow}\varphi_n\otimes\phi_m\\
&\stackrel{(\ZZ_{\rm e}\times \NN)+ 
(\ZZ_{\rm o}\times \NN)} 
{\longrightarrow}
\aaa{(\varphi_n\otimes\phi_m)\oplus 0& n={\rm even}\\
0\oplus (\varphi_n\otimes\phi_m)& n={\rm odd}.}
\end{align*}
The right-hand side is mapped as follows. 
\begin{align*}
&\stackrel{\rho\otimes\one}{\longrightarrow}
\aaa{(\varphi_{n/2}\otimes\phi_m)\oplus 0& n={\rm even}\\
{0}\oplus (\varphi_{(n-1)/2}\otimes\phi_m)& n={\rm odd}}
\stackrel{\tau} 
{\longrightarrow}
\aaa{\displaystyle
\binom{\varphi_{n/2}\otimes\phi_m}{0}& n={\rm even}\\
\ \\ 
\displaystyle \binom{0}{\varphi_{(n-1)/2}\otimes\phi_m}& n={\rm odd}}\\
&\stackrel{J}{\longrightarrow}
\aaa{\varphi_{n/2}\otimes\binom{\phi_m}{0}& n={\rm even}\\
\varphi_{(n-1)/2}\otimes\binom{0}{\phi_m}& n={\rm odd}}
\stackrel{\one\otimes\kappa}{\longrightarrow}
\aaa{\varphi_{n/2}\otimes{\varphi_m}& n={\rm even}\\
\varphi_{(n-1)/2}\otimes{\varphi_{-m-1}}& n={\rm odd}.}
\end{align*}
Therefore the lemma is proved.  
\qed

For example, $\phi_3 \otimes \phi_5$ is mapped to 
$\varphi_{-1} \otimes \varphi_{3}$, and 
$\phi_3 \otimes \phi_4$ is mapped to 
$\varphi_{-1} \otimes \varphi_{-4}$, etc. %See Figure \ref{zz}. 
Let 
$\tN^\rho=
(\rho\otimes \one) \tN^u (\rho^{-1} \otimes\one)$ be the total number operator in 
$(\lz\otimes\ln)\oplus(\lz\otimes \ln)$. 
The next lemma can be immediately proved.  
\begin{lemma}
We have 
\begin{align*}\tN^\rho
=
(2|N|\otimes \one +2\one\otimes \sM)\oplus 0+
0\oplus (|2N+\one|\otimes \one +2\one\otimes \sM).\end{align*}
\end{lemma}
The total number operator 
$\tN^\rho$ is transformed again to that on $\lz\otimes\lz$ as follows. 
Let 
$\tN^U=U\tN^\rho U^{-1}$. 
\begin{lemma}\label{50}
We have 
\begin{align*}
\tN^U
=2(|N|\otimes \one +\one\otimes N)(\one\otimes P _{[0,\infty)})+
(|2N+\one|\otimes\one+\one\otimes 2(|N|-\one)(\one\otimes P _{(-\infty, -1]}).
\end{align*}
\end{lemma}
\proof
The proof is similar to that of Lemma \ref{49}. 
\qed
The operator $N$ is the relative number operator on $\lz\otimes\lz$. 
Therefore 
$N\varphi_n\otimes \varphi_m=n\varphi_n\otimes \varphi_m$. 
On the other hand 
$\tN^U$ is the total number operator on $\lz\otimes\lz$. 
One can count the number of particles of 
$\varphi_{n} \otimes \varphi_m$ by 
$\tN^U$. 
\begin{lemma}\label{aabb}
We have 
\begin{align*}
\tN^U \varphi_{n} \otimes \varphi_m=
\aaa{(2|n|+2m)\varphi_{n} \otimes \varphi_m & m\geq0,\\
(|2n+1|-2(m+1))\varphi_{n} \otimes \varphi_m& m\leq -1.}\end{align*}
I.e.,
\begin{align}\label{AB}
\tN^U \varphi_{n} \otimes \varphi_m=
\aaa{
2(n+m)\varphi_{n} \otimes \varphi_m & n\geq0, m\geq0,\\
2(-n+m)\varphi_{n} \otimes \varphi_m & n\leq -1, m\geq0,\\
(2(n-m)-1)\varphi_{n} \otimes \varphi_m& n\geq0, m\leq -1,\\
(-2(n+m+1)-1)\varphi_{n} \otimes \varphi_m& n\leq-1, m\leq -1.
}
\end{align}
\end{lemma}
\proof
The proof is straightforward. We omit it. 
\qed
%By Lemma \ref{aabb} we see that 

\section{JJ-Hamiltonian $H_{S^1}$ on $\cH_{S^1}$}
\label{s4}
\subsection{Representation on $\cH_{S^1}$}
We shall represent $\jj $ on $\cH_{S^1}$ in this section. 
By the Fourier transform $F$ we can see that 
$\lz\cong L^2(S^1)$. 
Here $F: \lz\to L^2(S^1)$ is given by 
for $a=(a_n)_{n\in\ZZ}\in\lz$ and $\psi\in L^2(S^1)$, 
\begin{align*}
&(Fa)(\theta)=\frac{1}{\sqrt{2\pi}}\sum_{n\in\ZZ} a_ne^{-in\theta},\quad \theta\in S^1,\\
&(F^{-1}\psi)(n)=\frac{1}{\sqrt{2\pi}}\int_{S^1}\psi(\theta)e^{+in\theta}\rd \theta,\quad n\in\ZZ.
\end{align*}
The Fourier transform $F$ serves as a unitary between $\lz$ and $L^2(S^1)$, and $F\varphi_n(\theta)=e^{in\theta}/\sqrt{2\pi}$. 
Define $\sF $ by 
\begin{align*}\sF =F\otimes F.\end{align*}
 Then $\{e^{in\theta}/\sqrt{2\pi}\}_{n\in\ZZ}$ is a complete orthonormal system of $L^2(S^1)$. 
 Under the identification $$\cH_{S^1}\cong L^2(S^1\times S^1)$$ we can identify $e^{in\theta_1}\otimes e^{im\theta_2}$ with $e^{in\theta_1}e^{im\theta_2}$. 
We denote the projection 
$ F P _M F^{-1}$ on $L^2(S^1)$ by the same symbol 
$ P _M$, i.e., 
$$ P _M\psi(\theta)=\frac{1}{2\pi}
\sum_{n\in M} 
\lk \int_{S^1}\psi(\theta)e^{+in\theta}\rd \theta\rk 
e^{-in\theta}.$$
We define the self-adjoint operator $H_{S^1}$ on 
$\cH_{S^1}$ by 
\begin{align}\label{D4}
H_{S^1}=
\frac{1}{2C}
\lk 
-2i\frac{\partial}{\partial \theta_1}+q
\rk^2\otimes P _{[0,\infty)}
+
\frac{1}{2C}
\lk 
-2i\frac{\partial}{\partial \theta_1}+\one+q
\rk^2\otimes P _{(-\infty,-1]}
-\alpha H_{S^1,T},
\end{align}
where 
\begin{align*}
H_{S^1,T}
=A_{\{0\}}\otimes P _{\{0\}}+A_{[1,\infty)}\otimes P _{[1,\infty)}+A_{\{-1\}}\otimes P _{\{-1\}}+A_{(-\infty,-2]}\otimes P _{(-\infty,-2]},
\end{align*}
with 
\begin{align*}
&A_{\{0\}}=e^{i(\theta_1+\theta_2)} P _{(-\infty,-1]}+e^{-i(\theta_1-\theta_2)} P _{[1,\infty)},\\
&A_{[1,\infty)}=e^{-i(\theta_1+\theta_2)} P _{(-\infty,0]}+e^{-i(\theta_1-\theta_2)} P _{[1,\infty)} 
+e^{i(\theta_1+\theta_2)} P _{(-\infty,-1]}+
e^{i(\theta_1-\theta_2)} P _{[0,\infty)},\\
&A_{\{-1\}}=e^{i(\theta_1-\theta_2)} P _{(-\infty,-2]}
+e^{i\theta_1} P _{\{-1\}}+
e^{-i\theta_1} P _{\{0\}}+
e^{-i(\theta_1+\theta_2)} P _{[1,\infty)},\\
&A_{(-\infty,-2]}=e^{-i(\theta_1-\theta_2)} P _{(-\infty,-1]}
+e^{-i\theta_1} P _{\{0\}}
+e^{-i(\theta_1+\theta_2)} P _{[1,\infty)} \\
&\hspace{2cm}+ e^{i(\theta_1-\theta_2)} P _{(-\infty,-2]}
+e^{i\theta_1} P _{\{-1\}}
+e^{i(\theta_1+\theta_2)} P _{[0,\infty)}.
\end{align*}
Let $e_n(\theta)=e^{in\theta}$. 
In the representations of $H_{S^1,T}$ above, 
$e_n P _\#\otimes e_m P _\#$ is expressed as 
$e^{in\theta_1+im\theta_2} P _\#\otimes P _\#$. 
Let $\cU:\ln\otimes\ln\to \cH_{S^1}$(Figure \ref{uu}) 
 be defined by 
\begin{align}
\label{U}
\cU=\sF \circ \sU.
\end{align}
Now we are in the position to mention the main theorem in this paper. 
\begin{theorem}[Representation on $\cH_{S^1}$]\label{main}
We have 
\begin{align*}\jj \cong \hh^{f}\cong \hh^{u} \cong 
\hh^{\rho}\cong \hh^{U} \cong H_{S^1}.\end{align*} 
In particular 
$\cU\jj\cU^{-1}=H_{S^1}$. 
\end{theorem}
\proof
The first equivalence is proved in Lemma~\ref{32}, 
the second in Lemma~\ref{42}, 
the third in Lemma~\ref{45}, 
and the fourth in Lemma~\ref{49}. 
We now prove the final equivalence.
Note that $\sF :\lz\otimes\lz\to \cH_{S^1}$ is 
a unitary. 
Since 
$F N F^{-1}=-i\frac{\partial}{\partial \theta}$ and 
$F A F^{-1}=e^{-i\theta}$, 
we see that by Lemma \ref{49}
\begin{align*}
&
\sF ( P _++ P _-) \sF ^{-1}\\
&= 
e^{-i(\theta_1-\theta_2)} P _{[1,\infty)}\otimes P _{[0,\infty)} 
+ e^{-i(\theta_1+\theta_2)} P _{(-\infty,0]}\otimes P _{[1,\infty)}\\
&\hspace{1cm}
+
e^{i(\theta_1-\theta_2)} P _{[0,\infty)}\otimes P _{[1,\infty)}
+e^{i(\theta_1+\theta_2)} P _{(-\infty,-1]}\otimes P _{[0,\infty)}\\
&\hspace{1cm}+e^{-i(\theta_1+\theta_2)} P _{[1,\infty)} \otimes P _{(-\infty,-1]}
+e^{-i\theta_1} P _{\{0\}} \otimes P _{(-\infty,-1]}
+ e^{-i(\theta_1-\theta_2)} P _{(-\infty,-1]}\otimes 
 P _{(-\infty,-2]}\\
&\hspace{1cm}+ e^{i(\theta_1+\theta_2)} P _{[0,\infty)}\otimes P _{(-\infty,-2]} +
e^{i\theta_1} P _{\{-1\}}\otimes P _{(-\infty,-1]}
+e^{i(\theta_1-\theta_2)} P _{(-\infty,-2]}\otimes P _{(-\infty,-1]}\\
&
=A_{\{0\}}\otimes P _{\{0\}}+A_{[1,\infty)}\otimes P _{[1,\infty)}+A_{\{-1\}}\otimes P _{\{-1\}}+A_{(-\infty,-2]}\otimes P _{(-\infty,-2]}. 
\end{align*}
Then the theorem is proved.  
\qed
By Theorem \ref{main} we obtain the following corollary: 
\begin{corollary}\label{main2}
Let us suppose that 
 $\psi_1\in P _{[1,\infty)}\otimes P _{[1,\infty)}\cH_{S^1}$,
 $\psi_2\in P _{(-\infty,-2]}\otimes P _{(-\infty,-2]}\cH_{S^1}$,
 $\psi_3\in P _{(-\infty,-1]}\otimes P _{[1,\infty)}\cH_{S^1}$ and 
 $\psi_4\in P _{[1,\infty)}\otimes P _{(-\infty,-2]}\cH_{S^1}$.  
\iffalse
\begin{align*}
\begin{array}{ll}
 \psi_1\in P _{[1,\infty)}\otimes P _{[1,\infty)}\cH_{S^1}, & 
 \psi_2\in P _{(-\infty,-2]}\otimes P _{(-\infty,-2]}\cH_{S^1},\\
 \psi_3\in P _{(-\infty,-1]}\otimes P _{[1,\infty)}\cH_{S^1}, &
 \psi_4\in P _{[1,\infty)}\otimes P _{(-\infty,-2]}\cH_{S^1}. 
\end{array}
\end{align*}
\fi
Then 
\begin{align}\label{m1}
&H_{S^1}\psi_1(\theta_1,\theta_2)
=\frac{1}{2C}
\lk 
-2i\frac{\partial}{\partial \theta_1}+q
\rk^2\psi_1(\theta_1,\theta_2)
-2\alpha \cos(\theta_1-\theta_2) \psi_1(\theta_1,\theta_2),\\
\label{m11}&H_{S^1}\psi_2(\theta_1,\theta_2)
=\frac{1}{2C}
\lk 
-2i\frac{\partial}{\partial \theta_1}+\one+q
\rk^2\psi_2(\theta_1,\theta_2)
-2\alpha \cos(\theta_1-\theta_2) \psi_2(\theta_1,\theta_2),\\ 
% \end{align}
%and
%\begin{align}
\label{m2}
&H_{S^1}\psi_3(\theta_1,\theta_2)
=\frac{1}{2C}
\lk 
-2i\frac{\partial}{\partial \theta_1}+q
\rk^2\psi_3(\theta_1,\theta_2)
-2\alpha \cos(\theta_1+\theta_2) \psi_3(\theta_1,\theta_2),\\
\label{m22}
&H_{S^1}\psi_4(\theta_1,\theta_2)
=\frac{1}{2C}
\lk 
-2i\frac{\partial}{\partial \theta_1}+\one+q
\rk^2\psi_4(\theta_1,\theta_2)
-2\alpha \cos(\theta_1+\theta_2) \psi_4(\theta_1,\theta_2), 
\end{align}
\end{corollary}
\proof
We prove \eqref{m1}. The other statements are similarly  proved. 
By Theorem \ref{main} and the assumption we see that 
\begin{align*}
H_{S^1,T}\psi_1&=A_{[1,\infty)}\otimes P_{[1,\infty)}\psi_1\\
&=
(e^{-i(\theta_1-\theta_2)}P_{[1,\infty)}+e^{i(\theta_1-\theta_2)}P_{[0,\infty)})
\otimes P_{[1,\infty)}\psi_1=2\cos(\theta_1-\theta_2)\psi_1.\end{align*} 
Then \eqref{m1} follows. 
\qed
\subsection{Symmetric JJ-Hamiltonian} 
The kinetic term of the JJ-Hamiltonian on $\cH_{S^1}$ 
involves only the derivative with respect to $\theta_1$, 
and no derivative with respect to $\theta_2$ appears. 
Since $-i\frac{\partial}{\partial \theta_1}$ corresponds to the relative number operator, 
it is evident from the definition of the JJ-Hamiltonian on 
$\ln\otimes \ln$ that no $-i\frac{\partial}{\partial \theta_2}$ arises. 
Motivated by this observation, let us consider, albeit in an artificial manner, 
a Hamiltonian whose kinetic term symmetrically involves 
both $-i\frac{\partial}{\partial \theta_1}$ and $-i\frac{\partial}{\partial \theta_2}$.
Let \begin{align*}N_\pm=\tN-|\sN|.
\end{align*}
Therefore 
%\begin{align*}N_\pm \phi_\alpha\otimes \phi_\beta=(\alpha+\beta-|\alpha-\beta|)\phi_\alpha\otimes \phi_\beta=\aaa {2\alpha \phi_\alpha\otimes \phi_\beta&\alpha\leq \beta,\\2\beta \phi_\alpha\otimes \phi_\beta& \alpha>\beta.}\end{align*}
\begin{align*}
&N_\pm \phi_{n+m}\otimes \phi_m=2m \phi_{n+m}\otimes \phi_m,\\
&N_\pm \phi_{m}\otimes \phi_{n+m}=2m \phi_{m}\otimes \phi_{n+m}
\end{align*}
for any $n\geq0$. 
We define 
\begin{align*}\jjj=\frac{1}{2C}\sN^2+\frac{1}{2C}N_\pm^2-\alpha H_T.\end{align*}
Here we set $q=0$. 
By Lemma \ref{50} we can see that 
\begin{align*}N_\pm ^U=\sU N_\pm \sU^{-1}=
2\one\otimes N P _{[0,\infty)}-
\one\otimes 2(N+\one)P _{(-\infty, -1]}\end{align*}
and 
\begin{align}
N_\pm ^U\varphi_n\otimes \varphi_m=\aaa{2m\varphi_n\otimes \varphi_m&m\geq0,\\
-2(m+1)\varphi_n\otimes \varphi_m&m\leq-1.}
\end{align}
Then $\jjj$ can be transformed to 
the operator of the form
\begin{align*}
\sU\jjj\sU^{-1}&=
\frac{2}{C}\lk
N^2\otimes \one +
\one\otimes N^2\rk \one\otimes P _{[0,\infty)}\\
&+
\frac{2}{C}\lk
\lk N+\half\one\rk^2\otimes \one+
\one\otimes (N+\one)^2\rk 
\one\otimes P _{(-\infty, -1]}-\alpha H_T^U.
\end{align*}
By the Fourier transform $\sF $, 
$\sU\jjj\sU^{-1}$ can be also transformed to 
the operator $H_{S^1,\rm sym}$ in $\cH_{S^1}$: 
\begin{align}
H_{S^1,\rm sym}&=
\frac{2}{C}\lk
\lk -i\frac{\partial}{\partial \theta_1}\rk^2\otimes\one
+\one\otimes \lk -i\frac{\partial}{\partial \theta_2}\rk^2 
\rk 
\one\otimes P _{[0,\infty)}\nonumber\\
&\label{sym}+
\frac{2}{C}\lk
\lk -i\frac{\partial}{\partial \theta_1}+\half\one\rk^2\otimes\one+
\one\otimes \lk -i\frac{\partial}{\partial \theta_2}+\one\rk^2\rk 
\one\otimes P _{(-\infty, -1]}-\alpha H_{S^1,T}.
\end{align}
Therefore we finally obtain the Hamiltonian symmetrically involving 
$-i\frac{\partial}{\partial \theta_1}$ and $-i\frac{\partial}{\partial \theta_2}$. 

\begin{remark}[Physical interpretations of $\theta_1$ and $\theta_2$]\label{np}
For $\psi \in L^2(S^1)$, the function 
\[
\phi(\theta) = \theta \, \psi(\theta), \quad \theta \in S^1,
\]
is not periodic, and hence 
$\phi \notin L^2(S^1)$. 
Therefore, multiplication by $\theta$ does not define an operator on $L^2(S^1)$. 
Nevertheless, in physics, $\theta_{1}$ is formally regarded as canonically conjugate to the relative number operator 
$\sN \cong -2i \frac{\partial}{\partial \theta_{1}}$. 
$\sN$ acts on the state associated with the lattice point 
$(m+n,m)$ or $(m,m+n)$ in the $\NN \times \NN$ graph of Figure~\ref{x}, 
yielding the eigenvalue $n$ or $-n$, respectively. 
In parallel, $\theta_{2}$ is formally regarded as canonically conjugate to 
$N_{\pm}\cong -2i \frac{\partial}{\partial \theta_{2}}$, 
where $N_{\pm}$ acts by assigning to the state corresponding to 
$(m+n,m)$ or $(m,m+n)$ the eigenvalue $m$.
\end{remark}

\begin{remark}[Conjugate operators of $-i \tfrac{\partial}{\partial \theta}$]   
A conjugate operator associated with $-i \tfrac{\partial}{\partial \theta}$ in $L^2(S^1)$ has been studied in \cite{gal02a, AM08}. In particular, \cite{HT24a, HT24b,HT25} investigate conjugate operators associated with $\sM$. See Appendix \ref{conj}. 
\end{remark}

\section{Fiber decomposition}\label{s5}
\subsection{Interference and the Mathieu operator}
In this section we discuss a fiber decomposition of $H_{S^1}$. 
We begin with the fiber decomposition of $\jj$. Let $\ell_k=\ov{\rm LH}\{\phi_n\otimes\phi_m\in\cH\mid n+m=k\}$. Then 
$\tN\Phi=k\Phi$ for any $\Phi\in \ell_k$. 
Hence 
$$\cH=\bigoplus_{k=0}^\infty \ell_k.$$
By $[\tN , \mm]=0$, $\jj $ is reduced by each $\ell_k$. 
Therefore 
we have the fiber decomposition:
$$\jj =\bigoplus_{k=0}^\infty \jj \!\restriction_{\ell_k}.$$
We shall transform the fiber decomposition onto 
$\cH_{S^1}$ below. 
\iffalse
\begin{figure}[t]
\centering 
\begin{tikzpicture}[scale=0.35]
 \def\K{7} % 範囲の数値。表示ラベルは ±k に固定

 % 軸
 \draw[very thick,->] (-\K-0.5,0) -- (\K+0.7,0) node[below] {$m$};
 \draw[very thick,->] (0,-\K-0.5) -- (0,\K+0.7) node[left] {$n$};

 % グリッド
 \draw[very thin,gray!40] (-\K,-\K) grid (\K,\K);

 % 格子点
 \foreach \i in {-\K,...,\K}
 \foreach \j in {-\K,...,\K}
 \fill (\i,\j) circle (1.2pt);

 % 原点
 \node[below left] at (0,0) {\small 0};

 % 端の目盛りだけ ±k と表示
 \draw (-\K,0.12) -- (-\K,-0.12) node[below,yshift=+2pt] {\small $-k$};
 \draw ( \K,0.12) -- ( \K,-0.12) node[below,yshift=+2pt] {\small $k$};
 \draw (0.12,-\K) -- (-0.12,-\K) node[left,xshift=4pt] {\small $-k$};
 \draw (0.12, \K) -- (-0.12, \K) node[left,xshift=4pt] {\small $k$};

 % ご指定の太線
 \draw[very thick,blue] (0,\K) -- (\K,0);
 \draw[very thick,blue] (-1,\K-1) -- (-\K,0);
 \draw[very thick,red] (-\K,-1) -- (-1,-\K);
 \draw[very thick,red] (\K-1,-1) -- (0,-\K);
\end{tikzpicture}
\caption{The blue line represents the locus of points with total number 
$2k$, 
while the red line corresponds to those with total number 
$2k-1$.}
\label{k}
\end{figure}
\noindent
\fi
Set the total number operator in $\cH_{S^1}$ by 
\begin{align*}N_{S^1}=\sF 
\tN^U \sF ^{-1}.\end{align*}
It is explicitly given by 
\begin{align*}
N_{S^1}&=
2\lk 
\left|-i\frac{\partial}{\partial\theta_1}\right|
\otimes \one +\one\otimes -i\frac{\partial}{\partial\theta_2}
\rk
(\one\otimes P _{[0,\infty)})\\
&+
\lk 
\left|-2i\frac{\partial}{\partial\theta_1}+\one
\right|
\otimes\one-\one\otimes 2
\lk-i\frac{\partial}{\partial\theta_2}
+\one\rk\rk
(\one\otimes P _{(-\infty, -1]}).
\end{align*}
Since $e^{in\theta_1}\otimes e^{im\theta_2}\cong 2\pi \varphi_n\otimes\varphi_m$, 
it can be seen by \eqref{AB} that 
\begin{align}
\label{eo}
N_{S^1}e^{in\theta_1}\otimes e^{im\theta_2}=
\aaa{
2(n+m)e^{in\theta_1}\otimes e^{im\theta_2}& n\geq0, m\geq0,\\
2(-n+m)e^{in\theta_1}\otimes e^{im\theta_2}& n<0, m\geq0,\\
(2(n-m)-1)e^{in\theta_1}\otimes e^{im\theta_2}& n\geq0, m<0,\\
(-2(n+m+1)-1)e^{in\theta_1}\otimes e^{im\theta_2}& n<0, m<0.}
\end{align}
For $k\geq0$, let
$$L_k=\overline{\rm LH}\{e^{in\theta_1}\otimes e^{im\theta_2}\in \cH_{S^1}\mid 
N_{S^1}e^{in\theta_1}\otimes e^{im\theta_2} =k e^{in\theta_1}\otimes e^{im\theta_2}\}.$$ 
By \eqref{eo} 
$L_{2k}$ consists of functions of the form $e^{in\theta_1}e^{im\theta_2}$ with $m\geq0$, 
while 
$L_{2k-1}$consists of functions of the form $e^{in\theta_1}e^{im\theta_2}$ with $m<0$. 
More precisely we can see that 
\begin{align*}
L_{2k}&=
\overline{\rm LH}
\{
e^{in\theta_1}\otimes e^{im\theta_2}\mid m\geq0, 
n+m=k \mbox{ for } n\geq 0 \mbox{ or } -n+m=k \mbox{ for } n\leq -1
\}\\
&=\overline{\rm LH}
\{
e^{\pm in\theta_1}\otimes e^{i(k-n)\theta_2}\mid 0\leq n\leq k\},\\
L_{2k-1}&=
\overline{\rm LH}
\{
e^{in\theta_1}\otimes e^{im\theta_2}\mid m<0,
n-m=k \mbox{ for } n\geq 0 \mbox{ or } -n-m=k+1 \mbox{ for } n\leq -1
\}\\
&=\overline{\rm LH}
\{
e^{+in\theta_1}\otimes e^{-i(k-n)\theta_2},e^{-i(n+1)\theta_1}\otimes e^{-i(k-n)\theta_2}\mid 0\leq n\leq k-1\}.
\end{align*}
%See Figure \ref{k}. 
We obtain the decomposition: 
\begin{align*}\cH_{S^1}=\bigoplus_{k=0}^\infty L_k.\end{align*}
\begin{lemma}
We have 
\begin{align*}H_{S^1}=\bigoplus_{k=0}^\infty H_{S^1}\!\restriction_{L_k}.\end{align*}
\end{lemma}
\proof
Since $[H_{S^1},N_{S^1}]=0$ and $L_k$ is the eigenspace of $N_{S^1}$, 
$H_{S^1}$ is reduced by each $L_k$. Then the lemma is proved. 
\qed
In the theorem below we examine the action of $H_{S^1}$ on each fiber $L_k$. 
We shall employ the identification $\cH_{S^1}\cong L^2(S^1\times S^1)$ without further notice. 
Accordingly 
we identify 
$e^{in\theta_1}\otimes e^{im\theta_2}$ with 
$e^{in\theta_1} e^{im\theta_2}$.

\begin{theorem}[Actions on $L_{2k}$]\label{2k}
Let $k\geq2$, $a_0,a_n^\pm\in\CC$ for $n=1,\ldots,k$ and 
$$\psi(\theta_1,\theta_2)=\sum_\pm\sum_{1\leq n\leq k} a_n^\pm e^{\pm in\theta_1} e^{i(k-n)\theta_2}+
a_0 e^{ik\theta_2}
\in L_{2k}.$$ 
Then 
\begin{align*}
H_{S^1,T}\psi&=
a_k^{-}e^{i(\theta_1+\theta_2)}e^{-ik\theta_1}
+
a_k^{+}e^{-i(\theta_1-\theta_2)}e^{+ik\theta_1}
+2\cos\theta_1 a_0e^{i(k-1)\theta_2}\\
&+2\cos(\theta_1+\theta_2)
\sum_{1\leq n\leq k-1} 
a_n^{-} e^{-in\theta_1} e^{i(k-n)\theta_2}
+
2\cos(\theta_1-\theta_2)
\sum_{1\leq n\leq k-1} 
a_n^{+} e^{+in\theta_1} e^{i(k-n)\theta_2}. 
\end{align*}
\end{theorem}
\proof
$\psi$ is decomposed as 
\begin{align*}
\psi(\theta_1,\theta_2)=\sum_\pm a_k^\pm e^{\pm ik\theta_1} +a_0 e^{ik\theta_2} +
\sum_\pm\sum_{1\leq n\leq k-1} 
a_n^\pm e^{\pm in\theta_1} e^{i(k-n)\theta_2}. 
\end{align*}
%By Corollary \ref{main2} we expect that $\psi$ does {\it not} satisfy equations like \eqref{m1}-\eqref{m22}. 
Since $H_{S^1,T}\psi=(A_{\{0\}}\otimes P_{\{0\}}+
A_{[1,\infty)}\otimes P_{[1,\infty)})\psi$
and 
\begin{align*}
&A_{\{0\}}\otimes P_{\{0\}}+
A_{[1,\infty)}\otimes P_{[1,\infty)}\\
&=
\lk
e^{i(\theta_1+\theta_2)} P _{(-\infty,-1]}+e^{-i(\theta_1-\theta_2)} P _{[1,\infty)}\rk
\otimes P_{\{0\}}\\
&+\lk
2\cos(\theta_1+\theta_2)P_{(-\infty,-1]}
+
2\cos(\theta_1-\theta_2)P_{[1,\infty)}
+
2\cos\theta_1 e^{-i\theta_2}P_{\{0\}}\rk\otimes P_{[1,\infty)}, 
\end{align*}
we have 
\begin{align*}
&H_{S^1,T} \sum_\pm a_k^\pm e^{\pm ik\theta_1} 
=
e^{i(\theta_1+\theta_2)}a_k^{-}e^{-ik\theta_1}
+
e^{-i(\theta_1-\theta_2)}a_k^{+}e^{+ik\theta_1},\\
&
H_{S^1,T}a_0 e^{ik\theta_2}=
2\cos\theta_1 e^{-i\theta_2}a_0e^{ik\theta_2},\\
&H_{S^1,T}
\sum_\pm \sum_{1\leq n\leq k-1} 
a_n e^{\pm in\theta_1}e^{i(k-n)\theta_2}\\
&=
2\cos(\theta_1+\theta_2)
\sum_{1\leq n\leq k-1} 
a_n^{-} e^{-in\theta_1} e^{i(k-n)\theta_2}
+
2\cos(\theta_1-\theta_2)
\sum_{1\leq n\leq k-1} 
a_n^{+} e^{+in\theta_1} e^{i(k-n)\theta_2}. 
\end{align*}
Then the theorem follows. 
\qed
By Theorem \ref{2k} it can be straightforwardly verified that $H_{S^1,T}\psi\in L_{2k}$. 
As a special case of Theorem \ref{2k} we obtain the following corollary. 
\begin{corollary}\label{k0}
Let $k\geq2$, $a_0,a_n^\pm\in\CC$ for $n=1,\ldots,k-1$, $a_k^{\pm}=0$ 
 and 
$$\psi(\theta_1,\theta_2)=\sum_\pm\sum_{1\leq n\leq k} a_n^\pm e^{\pm in\theta_1} e^{i(k-n)\theta_2}+
a_0 e^{ik\theta_2}
\in L_{2k}.$$ 
Then 
\begin{align*}
&H_{S^1}\psi=\frac{1}{2C}\lk -2i\frac{\partial}{\partial \theta_1}+q\rk^2\psi
-2\alpha
\cos\theta_1 a_0e^{i(k-1)\theta_2}\\
&-2\alpha\lk \cos(\theta_1+\theta_2)
\sum_{1\leq n\leq k-1} 
a_n^{-} e^{-in\theta_1} e^{i(k-n)\theta_2}
+
\cos(\theta_1-\theta_2)
\sum_{1\leq n\leq k-1} 
a_n^{+} e^{+in\theta_1} e^{i(k-n)\theta_2}\rk.
\end{align*}
\end{corollary}

In the case of $L_{2k-1}$ one can obtain a similar result. 
\begin{theorem}[Actions on $L_{2k-1}$]\label{2k-1}
Let $k\geq2$, $a_n^\pm\in\CC$ for $n=0,1,\ldots,k-1$ and 
$$\psi(\theta_1,\theta_2)=\sum_{0\leq n\leq k-1} 
(a_n^{+} e^{+ in\theta_1} +
a_n^{-} e^{- i(n+1)\theta_1})
e^{-i(k-n)\theta_2}
\in L_{2k-1}.$$ 
Then  
\begin{align*}
H_{S^1,T}\psi&=
(a_{k-1}^{+}e^{-i(\theta_1+\theta_2)}
 e^{+ i(k-1)\theta_1} +
a_{k-1}^{-}
e^{i(\theta_1-\theta_2)}
 e^{- ik\theta_1})
e^{-i\theta_2}\\
&+2\cos(\theta_1+\theta_2)
\sum_{0\leq n\leq k-2} 
a_n^{+} e^{+ in\theta_1} e^{-i(k-n)\theta_2}\\
&+
2\cos(\theta_1-\theta_2)
\sum_{0\leq n\leq k-2} a_n^{-} e^{- i(n+1)\theta_1}
e^{-i(k-n)\theta_2}. 
\end{align*}
\end{theorem}
\proof
The proof is similar to that of Theorem \ref{2k}. 
$\psi$ is decomposed as 
\begin{align*}
\psi(\theta_1,\theta_2)
=\sum_{0\leq n\leq k-2} 
(a_n^{+} e^{+ in\theta_1} +
a_n^{-} e^{- i(n+1)\theta_1})
e^{-i(k-n)\theta_2}+
(a_{k-1}^{+} e^{+ i(k-1)\theta_1} +
a_{k-1}^{-} e^{- ik\theta_1})
e^{-i\theta_2}. 
\end{align*}
Since 
$
H_{S^1,T}\psi
=(A_{\{-1\}}\otimes P _{\{-1\}}+A_{(-\infty,-2]}\otimes P _{(-\infty,-2]})\psi$ 
and
\begin{align*}
&A_{\{-1\}}=e^{i(\theta_1-\theta_2)} P _{(-\infty,-2]}
+e^{i\theta_1} P _{\{-1\}}+
e^{-i\theta_1} P _{\{0\}}+
e^{-i(\theta_1+\theta_2)} P _{[1,\infty)},\\
&A_{(-\infty,-2]}=(e^{-i(\theta_1-\theta_2)} +e^{i\theta_1} )P _{\{-1\}}
+(e^{-i\theta_1} +e^{i(\theta_1+\theta_2)} )P _{\{0\}}\\
&\hspace{2cm}+2\cos(\theta_1+\theta_2) P _{[1,\infty)} 
+ 2\cos(\theta_1-\theta_2) P _{(-\infty,-2]}, 
\end{align*}
we have 
\begin{align*}
&H_{S^1,T}(a_{k-1}^{+} e^{+ i(k-1)\theta_1} +
a_{k-1}^{-} e^{- ik\theta_1})
e^{-i\theta_2}\\
&=
(e^{-i(\theta_1+\theta_2)}
a_{k-1}^{+} e^{+ i(k-1)\theta_1} +
e^{i(\theta_1-\theta_2)}
a_{k-1}^{-} e^{- ik\theta_1})
e^{-i\theta_2},\\
&H_{S^1,T}
\sum_{0\leq n\leq k-2} 
(a_n^{+} e^{+ in\theta_1} +
a_n^{-} e^{- i(n+1)\theta_1})
e^{-i(k-n)\theta_2}\\ 
&=
\sum_{0\leq n\leq k-2} 
(2\cos(\theta_1+\theta_2)a_n^{+} e^{+ in\theta_1} +
2\cos(\theta_1-\theta_2)a_n^{-} e^{- i(n+1)\theta_1})
e^{-i(k-n)\theta_2}. 
\end{align*}
Then the theorem follows. 
\qed
As a special case of Theorem \ref{2k-1} we obtain the following corollary. 
\begin{corollary}\label{k-10}
Let $k\geq2$, $a_n^\pm\in\CC$ for $n=0,1,\ldots,k-2$, $a_{k-1}^{\pm}=0$ 
 and 
$$\psi(\theta_1,\theta_2)=\sum_{0\leq n\leq k-1} 
(a_n^{+} e^{+ in\theta_1} +
a_n^{-} e^{- i(n+1)\theta_1})
e^{-i(k-n)\theta_2}
\in L_{2k-1}.$$ 
Then 
\begin{align*}
&H_{S^1}\psi=\frac{1}{2C}\lk -2i\frac{\partial}{\partial \theta_1}+1+q\rk^2\psi\\ 
&-2\alpha
\lk
\cos(\theta_1+\theta_2)
\sum_{0\leq n\leq k-2} 
a_n^{+} e^{+ in\theta_1}e^{-i(k-n)\theta_2} +
\cos(\theta_1-\theta_2)
\sum_{0\leq n\leq k-2} 
a_n^{-} e^{- i(n+1)\theta_1}
e^{-i(k-n)\theta_2}\rk. 
\end{align*}
\end{corollary}

We derive a Mathieu operator \eqref{ma1} on the fiber with fixed particle number below.
\begin{corollary}[Mathieu operator]\label{ma}
Let $\psi$ be in Corollary \ref{k0} and $q=0$, or
$\psi$ be in Corollary \ref{k-10} and $q=-1$. 
Then 
\begin{align}\label{ma1}
(H_{S^1}\psi)(\theta,0)
=\frac{2}{C}\lk -i\frac{\partial}{\partial \theta}\rk^2\psi(\theta,0)
-2\alpha\cos\theta \psi(\theta,0).
\end{align}
\end{corollary}
\proof
This follows from Corollaries \ref{k0} and \ref{k-10}.  
\qed

\subsection{Discussion on no interference}
In Corollary \ref{k0} it is assumed that 
$a_k^{\pm}=0$ for 
$\psi(\theta_1,\theta_2)\in L_{2k}$ 
and 
in Corollary \ref{k-10} 
$a_{k-1}^{\pm}=0$ is assumed for 
$\psi(\theta_1,\theta_2)\in L_{2k-1}$.
Let us now unravel the underlying meaning. 
Suppose that $a_k^{\pm} \neq 0$ while all other coefficients vanish for $\psi$ in Theorem~\ref{2k}. 
Then we obtain 
\begin{align*}
 \psi_0 = a_k^{+} e^{i k \theta_1} + a_k^{-} e^{-i k \theta_1}  \in  L_{2k}.
\end{align*}
On the other hand, 
if $a_{k-1}^{\pm} \neq 0$ while the remaining coefficients vanish for $\psi$ in Theorem~\ref{2k-1}, then 
\begin{align*}
 \psi_1 = \bigl(a_{k-1}^{+} e^{i (k-1) \theta_1} + a_{k-1}^{-} e^{-i k \theta_1}\bigr) e^{-i \theta_2}  \in  L_{2k-1}.
\end{align*}
A direct computation shows that 
\begin{align*}
 H_{S^1,T} \psi_0(\theta_1,\theta_2)
 = e^{i(\theta_1+\theta_2)} a_k^{-} e^{-i k \theta_1}
 + e^{-i(\theta_1-\theta_2)} a_k^{+} e^{i k \theta_1},
\end{align*}
which implies in particular that 
\begin{align*}
 H_{S^1,T} \psi_0(\theta_1,0)  \neq  \cos\theta_1 \psi_0(\theta_1,0).
\end{align*}
Similarly we can see that 
\begin{align*}
 H_{S^1,T} \psi_1(\theta_1,0)  \neq  \cos\theta_1 \psi_1(\theta_1,0).
\end{align*}
By Lemma~\ref{ab} it is proved that 
%\iffalse
\begin{align*}
 \sU \phi_\alpha \otimes \phi_\beta =
\aaa{
 \varphi_{n/2} \otimes \varphi_m, & n\  \text{\rm even},\\
 \varphi_{(n-1)/2} \otimes \varphi_{-m-1}, & n\   \text{\rm odd},
}
\end{align*}
%\fi
where $n = \alpha-\beta$ and $m = \min\{\alpha,\beta\}$, and where $\sU$ is defined in~\eqref{UU}. 
By Lemma \ref{aabb} it is also proved that 
for even $n$, 
\begin{align*}
\tN^U \varphi_{n/2} \otimes \varphi_m=\aaa{(|n|+2m)\varphi_{n/2} \otimes \varphi_m & m\geq0,\\
(|n+1|-2(m+1))\varphi_{n/2} \otimes \varphi_m& m\leq -1,}\end{align*}
for odd $n$, 
\begin{align*}\tN^U \varphi_{(n-1)/2} \otimes \varphi_{-m-1}=\aaa{(|n|+2m)\varphi_{(n-1)/2} \otimes \varphi_{-m-1} & m\geq0,\\
(|n-1|-2(m+1))\varphi_{(n-1)/2} \otimes \varphi_{-m-1}& m\leq -1.}\end{align*}
Consequently, 
$e^{i k \theta_1} e^{i0 \theta_2}\cong \varphi_k\otimes\varphi_0$ and $e^{-i k \theta_1} e^{i0 \theta_2}\cong 
\varphi_{-k}\otimes\varphi_0$
appearing in $\psi_0$ correspond to 
$\phi_{2k} \otimes \phi_0$ and $\phi_0 \otimes \phi_{2k}$ in $\ln \otimes \ln $, respectively. 
Similarly, 
$e^{i (k-1) \theta_1} e^{-i \theta_2}\cong \varphi_{k-1}\otimes\varphi_{-1}$ and 
$e^{-i k \theta_1} e^{-i \theta_2}\cong \varphi_{-k}\otimes\varphi_{-1}$
appearing in $\psi_1$ correspond to 
$\phi_{2k-1} \otimes \phi_0$ and $\phi_0 \otimes \phi_{2k-1}$, respectively. 
Notably, each of the vectors 
$\phi_{2k} \otimes \phi_0$, 
$\phi_0 \otimes \phi_{2k}$, 
$\phi_{2k-1} \otimes \phi_0$, 
and $\phi_0 \otimes \phi_{2k-1}$
represents a configuration in which all particles are localized on one side. 
Hence, particle transfer can occur only in a single direction. 
As a consequence, {\it no interference arises in the tunneling process}. 
Hence no Mathieu operator appears for $\psi_0$ and $\psi_1$.

\subsection{Spectrum of $\jj$}
$L_{2k}$ and $L_{2k+1}$ are the finite dimensional subspace of $\cH_{S^1}$ 
and $H_{S^1}$ can be reduced by these spaces. 
The matrix representation of $H_{S^1}\restriction_{L_\#}$ can be easily given. 
We choose a base 
\begin{align*}
\{e_k,e_{k-1},\ldots, e_0, e_{-1},e_{-2},\ldots,e_{-k}\}
\end{align*}
of $L_{2k}$, where 
$e_n=e^{in\theta_1}e^{i(k-|n|)\theta_2}$. 
By the proof of Theorem \ref{2k} 
we can see that 
the matrix representation of $H_{S^1}\restriction_{L_{2k}}$ under the base 
above 
is give by 
\begin{align*}
H_{S^1}\restriction_{L_{2k}}=M_{2k}=
\lk
\begin{smallmatrix}
(2k-q)^2&-\alpha&0&0&0&0&\ldots &0\\
-\alpha&(2k-2-q)^2&-\alpha&0&0&0&\ldots &0 \\
0& -\alpha&(2k-4-q)^2&-\alpha&0&0&\ldots &0 \\
0& 0& -\alpha&(2k-6-q)^2&-\alpha&0&\ldots &0 \\
0& 0& 0& -\alpha&(2k-8-q)^2&-\alpha&\ldots &0 \\
%\vdots& \vdots& \vdots& \vdots&1&(k-5-q)^2&\vdots &\vdots\\
\vdots& \vdots& \vdots& \vdots&-\alpha&\vdots&\vdots &\vdots\\
\vdots& \vdots& \vdots& \vdots&\vdots&\vdots&\vdots &-\alpha\\
\vdots& \vdots& \vdots& \vdots&\vdots&\vdots&-\alpha &(-2k-q)^2
\end{smallmatrix}
\rk
\end{align*}
Similarly in the case of $L_{2k-1}$ 
we choose a base 
\begin{align*}
&\{e_k,e_{k-1},\ldots, e_0, e_{-1},e_{-2},\ldots,e_{-k}\}\\
&=
\{e^{i(k-1)\theta_1}e^{-i\theta_2},
e^{i(k-2)\theta_1}e^{-2i\theta_2},\ldots, e^{-ik\theta_2},
e^{-i\theta_1}e^{-ik\theta_2},
e^{-2i\theta_1}e^{-i(k-1)\theta_2},
\ldots,
e^{-ik\theta_1}e^{-i\theta_2}\}.
\end{align*}
By Theorem \ref{2k-1} we can see that 
the matrix representation of $H_{S^1}\restriction_{L_{2k-1}}$ under the base 
bove 
is give by 
\begin{align*}
H_{S^1}\restriction_{L_{2k-1}}=M_{2k-1}=
\lk
\begin{smallmatrix}
(2k-1-q)^2&-\alpha&0&0&0&0&\ldots &0\\
-\alpha&(2k-3-q)^2&-\alpha&0&0&0&\ldots &0 \\
0& -\alpha&(2k-5-q)^2&-\alpha&0&0&\ldots &0 \\
0& 0& -\alpha&(2k-7-q)^2&-\alpha&0&\ldots &0 \\
0& 0& 0& -\alpha&(2k-9-q)^2&-\alpha&\ldots &0 \\
%\vdots& \vdots& \vdots& \vdots&1&(k-5-q)^2&\vdots &\vdots\\
\vdots& \vdots& \vdots& \vdots&-\alpha&\vdots&\vdots &\vdots\\
\vdots& \vdots& \vdots& \vdots&\vdots&\vdots&\vdots &-\alpha\\
\vdots& \vdots& \vdots& \vdots&\vdots&\vdots&-\alpha &(-2k+1-q)^2
\end{smallmatrix}
\rk
\end{align*}
\begin{theorem}[Spectrum of $H_{S^1}$]
The spectrum of $H_{S^1}$ is given by
\begin{align*}\sigma(H_{S^1})=\overline{\bigcup_{k=0}^\infty \sigma(M_k)}\end{align*}
and 
\begin{align*}\sigma_p(H_{S^1})\subset {\bigcup_{k=0}^\infty \sigma(M_k)}.\end{align*}
\end{theorem}
\proof
By the matrix representations above we can see that 
$H_{S^1}=\bigoplus_{k=0}^\infty M_k$. 
Then the theorem is proved. 
\qed

\section{Josephson current and Fraunhofer pattern}\label{7}

\subsection{Josephson current}
The Josephson effect is one of the most striking manifestations of macroscopic quantum coherence. When two superconductors are weakly coupled through a thin insulating barrier, 
a supercurrent can flow across the junction without any applied voltage. 
This current, known as the Josephson current, arises from the quantum mechanical tunneling of Cooper pairs and is governed by a simple but fundamental relation: 
it depends sinusoidally on the  phase shift  between the superconducting order parameters on both sides of the junction. The Josephson current thus provides a direct link between phase coherence in superconductors and measurable electrical transport.
%, and serves as the cornerstone of a wide range of quantum devices, from superconducting qubits to ultra\UTF{2013}sensitive magnetometers.

In this section we study the magnetic JJ-Hamiltonian $ \mm $.
We begin with formulating a rigorous definition of the Josephson current and proceed
to analyze its magnetic response, elucidating how the current depends on the
magnetic field in the framework developed below.
\begin{lemma}
The operator $ \mm $ can also be represented on $ \lz\otimes\lz$ as 
\begin{align}\label{D7}
U^{-1} \mm U 
= H_{+}\otimes P_{[0,\infty)}
+ H_{-}\otimes P_{(-\infty,-1]}
- \alpha\bigl(P_{+}(\Phi) + P_{-}(\Phi)\bigr),
\end{align}
where 
\begin{align*}
 P _+(\Phi)= &e^{i\Phi}\lkk
 A P _{[1,\infty)}\otimes A^\ast P _{[0,\infty)} 
+ A P _{(-\infty,0]}\otimes A P _{[1,\infty)}\rkk
\\
&+
e^{-i\Phi}\lkk
A^{\ast} P _{[0,\infty)}\otimes A P _{[1,\infty)}
+A^{\ast} P _{(-\infty,-1]}\otimes A^\ast P _{[0,\infty)}\rkk,\\
 P _-(\Phi)=&
e^{i\Phi}A \lkk 
P _{[1,\infty)} \otimes A P _{(-\infty,-1]}
+A P _{\{0\}} \otimes P _{(-\infty,-1]}+ A P _{(-\infty,-1]}\otimes A^\ast P _{(-\infty,-2]}\rkk\\
&+
e^{-i\Phi}\lkk
A^{\ast} P _{[0,\infty)}\otimes A^\ast P _{(-\infty,-2]} +
A^{\ast} P _{\{-1\}}\otimes P _{(-\infty,-1]}
+A^{\ast} P _{(-\infty,-2]}\otimes A P _{(-\infty,-1]}\rkk.
\end{align*}
\end{lemma}
\proof
The proof is parallel to the representation of $ \jj $ on 
$ \lz \otimes \lz $ given in \eqref{D6}. Then we omit it. 
\qed
We define the self-adjoint operator $ H_{S^1}(\Phi) $ on 
$ \cH_{S^1} $ by 
\begin{align}\label{D8}
H_{S^1}(\Phi)=
\frac{1}{2C}
\left(-2i\frac{\partial}{\partial \theta_1}+q\right)^2 \otimes P_{[0,\infty)}
+
\frac{1}{2C}
\left(-2i\frac{\partial}{\partial \theta_1}+\one+q\right)^2 \otimes P_{(-\infty,-1]}
- \alpha H_{S^1,T}(\Phi),
\end{align}
where 
\begin{align*}
H_{S^1,T}(\Phi)
=B_{\{0\}}\otimes P _{\{0\}}+B_{[1,\infty)}\otimes P _{[1,\infty)}+B_{\{-1\}}\otimes P _{\{-1\}}+B_{(-\infty,-2]}\otimes P _{(-\infty,-2]},
\end{align*}
with 
\begin{align*}
&B_{\{0\}}=e^{i(\theta_1-\Phi+\theta_2)} P _{(-\infty,-1]}+e^{-i(\theta_1-\Phi-\theta_2)} P _{[1,\infty)},\\
&B_{[1,\infty)}=e^{-i(\theta_1-\Phi+\theta_2)} P _{(-\infty,0]}+e^{-i(\theta_1-\Phi-\theta_2)} P _{[1,\infty)} 
+e^{i(\theta_1-\Phi+\theta_2)} P _{(-\infty,-1]}+
e^{i(\theta_1-\Phi-\theta_2)} P _{[0,\infty)},\\
&B_{\{-1\}}=e^{i(\theta_1-\Phi-\theta_2)} P _{(-\infty,-2]}
+e^{i\theta_1-\Phi} P _{\{-1\}}+
e^{-i\theta_1-\Phi} P _{\{0\}}+
e^{-i(\theta_1-\Phi+\theta_2)} P _{[1,\infty)},\\
&B_{(-\infty,-2]}=e^{-i(\theta_1-\Phi-\theta_2)} P _{(-\infty,-1]}
+e^{-i\theta_1-\Phi} P _{\{0\}}
+e^{-i(\theta_1-\Phi+\theta_2)} P _{[1,\infty)} \\
&\hspace{2cm}+ e^{i(\theta_1-\Phi-\theta_2)} P _{(-\infty,-2]}
+e^{i\theta_1-\Phi} P _{\{-1\}}
+e^{i(\theta_1-\Phi+\theta_2)} P _{[0,\infty)}.
\end{align*}

\begin{lemma}\label{gauge2}
It follows that 
\begin{align}\label{v1}
e^{-\Phi \frac{\partial}{\partial \theta_1}}
H_{S^1}
e^{\Phi \frac{\partial}{\partial \theta_1}}
= H_{S^1}(\Phi)
\end{align}
and then 
\begin{align}\label{v2}
\s(\jj)
= \s(H_{S^1}(\Phi))
\end{align}
for any $\Phi\in\RR$. 
\end{lemma}
\proof
\kak{v1}  follows from 
Proposition \ref{gauge} and $-2i\frac{\partial}{\partial \theta_1}\cong \sN$, 
and \kak{v2} follows from \kak{v1}. 
\qed
In the representation on $ \cH_{S^1} $, 
the Josephson current $I_{\rm JJ}(\Phi)$ can be expressed as 
\begin{align*}
I_{S^1}(\Phi) = \Bigl[\frac{\partial}{\partial \theta_1}, H_{S^1}(\Phi)\Bigr].
\end{align*}
It is shown in Lemma \ref{bounded} that 
$I_{S^1}(\Phi)$ is  a bounded operator for any $\Phi\in\RR$. 
Let us set 
\[p_1=P_{(-\infty, -2]},\quad
p_2=P_{\{-1\}},\quad 
p_3=P_{\{0\}},\quad 
p_4=P_{[1,\infty)}.\] 
We have $p_ip_j=0$ for $i\neq j$ and
\[p_1+p_2+p_3+p_4=\one.\] 
Hence $\mathcal{H}_{S^1}$ can be decomposed into 16 mutually orthogonal subspaces:  
\[\cH_{S^1}=\bigoplus_{1\leq i,j\leq 4} p_i\otimes p_j \cH_{S^1}.\] 
\begin{theorem}[Decomposition of the Josephson current]
It follows that 
\begin{align}\label{c7}
I_{S^1}(\Phi)=-\alpha\bigoplus_{1\leq i,j\leq 4} K_{ij}p_i\otimes p_j.
\end{align}
Here $K_{ij}$ is a multiplication operator 
given in Figure \ref{kij}. 
In particular 
$I_{S^1}(\Phi)$ is reduced by $p_i\otimes p_j\cH_{S^1}$ for each $1\leq i,j\leq 4$.  
\renewcommand{\arraystretch}{1.7}
\begin{figure}[h]
\centering 
\resizebox{1\textwidth}{!}{%
\begin{tabular}{c|c|c|c|c}
$p_i\otimes p_j\cH_{S^1}$ &$p_1$ &$p_2$ &$p_3$ &$p_4$ \\ \hline
$p_1$ &$-2\sin(\theta_1-\Phi-\theta_2)$ & $ie^{i(\theta_1-\Phi-\theta_2)}$&$ie^{i(\theta_1-\Phi+\theta_2)}$ & $-2\sin (\theta_1-\Phi+\theta_2)$\\ \hline
$p_2$ &$-ie^{-i(\theta_1-\Phi-\theta_2)}+ie^{i(\theta_1-\Phi)}$ &$ie^{i(\theta_1-\Phi)}$ & $ie^{i(\theta_1-\Phi+\theta_2)}$& $-2\sin (\theta_1-\Phi+\theta_2)$\\ \hline
$p_3$ &$ -ie^{-i(\theta_1-\Phi)}+ie^{i(\theta_1-\Phi+\theta_2)}$ &$ie^{-i(\theta_1-\Phi)}$ & $0$&$ -ie^{-i(\theta_1-\Phi+\theta_2)}+ie^{i(\theta_1-\Phi-\theta_2)}$ \\ \hline
$p_4$ &$-2\sin(\theta_1-\Phi+\theta_2)$ &$-ie^{-i(\theta_1-\Phi+\theta_2)}$ &$-ie^{-i(\theta_1-\Phi-\theta_2)} $ & $-2\sin (\theta_1-\Phi-\theta_2)$
\end{tabular}
}\caption{$K_{ij}$: action of $I_{S^1}(\Phi)$ on $p_i\otimes p_j \cH_{S^1}$}
\label{kij}
\end{figure}\end{theorem}
\proof
Since 
\begin{align*}
I_{S^1}(\Phi) 
= -\alpha \Bigl[\frac{\partial}{\partial \theta_1}, H_{S^1,T}(\Phi)\Bigr],
\end{align*}
we obtain 
\begin{align*}
I_{S^1}(\Phi) 
=-\alpha 
\lkk 
C_{\{0\}} \otimes P_{\{0\}}
+ C_{[1,\infty)} \otimes P_{[1,\infty)}
 + 
C_{\{-1\}} \otimes P_{\{-1\}}
+ C_{(-\infty,-2]} \otimes P_{(-\infty,-2]}
\rkk, 
\end{align*}
where 
$C_\#$ is the derivative of $B_\#$ with respect to $\theta_1$: 
\begin{align*}
&C_{\{0\}}=i\lkk e^{i(\theta_1-\Phi+\theta_2)} P _{(-\infty,-1]}-e^{-i(\theta_1-\Phi-\theta_2)} P _{[1,\infty)}\rkk,\\
&C_{[1,\infty)}=i\lkk \!-e^{-i(\theta_1-\Phi+\theta_2)} P _{(-\infty,0]}-e^{-i(\theta_1-\Phi-\theta_2)} P _{[1,\infty)} 
\!+e^{i(\theta_1-\Phi+\theta_2)} P _{(-\infty,-1]}
\!+
e^{i(\theta_1-\Phi-\theta_2)} P _{[0,\infty)}\rkk\!,\\
&C_{\{-1\}}=i\lkk
e^{i(\theta_1-\Phi-\theta_2)} P _{(-\infty,-2]}
+e^{i(\theta_1-\Phi)} P _{\{-1\}}-
e^{-i(\theta_1-\Phi)} P _{\{0\}}
-e^{-i(\theta_1-\Phi+\theta_2)} P _{[1,\infty)}\rkk ,\\
&C_{(-\infty,-2]}=i\lkk -e^{-i(\theta_1-\Phi-\theta_2)} P _{(-\infty,-1]}
-e^{-i(\theta_1-\Phi)} P _{\{0\}}
-e^{-i(\theta_1-\Phi+\theta_2)} P _{[1,\infty)} \right.\\
&\left.\hspace{2cm}+ e^{i(\theta_1-\Phi-\theta_2)} P _{(-\infty,-2]}
+e^{i(\theta_1-\Phi)} P _{\{-1\}}
+e^{i(\theta_1-\Phi+\theta_2)} P _{[0,\infty)}\rkk.
\end{align*}
Then the theorem follows. 
\qed

\subsection{Sinusoidal phase dependence}
A central feature of the Josephson effect is the emergence of a supercurrent 
that flows across a junction without any applied voltage.  
This current arises from the coherent tunneling of Cooper pairs and is governed by 
a fundamental phase relation between the macroscopic wave function of the Cooper pairs  
on both sides of the junction.  
In its simplest form, the Josephson current depends sinusoidally on the phase difference, 
providing a direct link between macroscopic phase coherence and measurable electrical transport.

We can see the action of the Josephson current on 
$L_{2k}$ and $L_{2k-1} $ exactly. 
\begin{corollary}[Josephson current on $L_{2k}$]\label{MAIN1}
Let $k\geq2$, $a_0,a_n^\pm\in\CC$ for $n=1,\ldots,k-1$ and 
\begin{align}\label{st1}
\psi(\theta_1,\theta_2)=\sum_\pm\sum_{1\leq n\leq k-1} a_n^\pm e^{\pm in\theta_1} e^{i(k-n)\theta_2}+
a_0 e^{ik\theta_2}
\in L_{2k}.
\end{align}
Then 
\begin{align}
(I_{S^1}(\Phi)\psi)(\theta_1,\theta_2)=&
2\alpha
\sin(\theta_1-\Phi)a_0e^{i(k-1)\theta_2}
+2\alpha
\sin(\theta_1-\Phi+\theta_2)
\sum_{1\leq n\leq k-1} 
a_n^{-} e^{-in\theta_1} e^{i(k-n)\theta_2}\nonumber \\
&\label{c1}+2\alpha
\sin(\theta_1-\Phi-\theta_2)
\sum_{1\leq n\leq k-1} 
a_n^{+} e^{+in\theta_1} e^{i(k-n)\theta_2}. 
\end{align}
In particular  
\begin{align}
(I_{S^1}(\Phi)\psi)(\theta_1,0)=&
2\alpha\sin(\theta_1-\Phi)
\lk
a_0
+
\sum_{1\leq n\leq k-1} 
a_n^{-} e^{-in\theta_1} 
+\sum_{1\leq n\leq k-1} 
a_n^{+} e^{+in\theta_1}\rk. 
\end{align}
\end{corollary}
\proof
The proof of \kak{c1} is similar to that of Theorem \ref{2k}. 
\qed
The Josephson current on $L_{2k-1}$ can be also computed. 
\begin{corollary}[Josephson current on $L_{2k-1}$]
\label{MAIN2}
Let $k\geq3$, $a_n^\pm\in\CC$ for $n=0,1,\ldots,k-2$ 
 and 
\begin{align}\label{st2}
\psi(\theta_1,\theta_2)=\sum_{0\leq n\leq k-2} 
(a_n^{+} e^{+ in\theta_1} +
a_n^{-} e^{- i(n+1)\theta_1})
e^{-i(k-n)\theta_2}
\in L_{2k-1}.\end{align}
Then 
\begin{align}
(I_{S^1}(\Phi)\psi)(\theta_1,\theta_2)=&
2\alpha
\sin(\theta_1-\Phi+\theta_2)
\sum_{0\leq n\leq k-2} 
a_n^{+} e^{+ in\theta_1}e^{-i(k-n)\theta_2} \nonumber \\
&\label{c3}+
2\alpha \sin(\theta_1-\Phi-\theta_2)
\sum_{0\leq n\leq k-2} 
a_n^{-} e^{- i(n+1)\theta_1}
e^{-i(k-n)\theta_2}. 
\end{align}
In particular 
\begin{align}
(I_{S^1}(\Phi)\psi)(\theta_1,0)=&
2\alpha
\sin(\theta_1-\Phi)
\lk
\sum_{0\leq n\leq k-2} 
a_n^{+} e^{+ in\theta_1}+
\sum_{0\leq n\leq k-2} 
a_n^{-} e^{- i(n+1)\theta_1}\rk.
\end{align}
\end{corollary}
\proof 
The proof is similar to that of Corollary \ref{MAIN1}. 
\qed

\subsection{Aharonov-Bohm effect and Josephson current}
The Aharonov-Bohm effect \cite{AB59} shows that in quantum mechanics, 
charged particles are influenced by vector potentials $A$ even in regions where the corresponding 
magnetic fields $\nabla\times A$ vanish. See Appendix \ref{app}. 
An electron beam encircling a confined  phase shift  acquires a measurable  phase shift , 
demonstrating the physical significance of vector potentials and the nonlocal nature of quantum theory.

In Lemma \ref{gauge2}  we show that 
$e^{-\Phi \frac{\partial}{\partial \theta_1}}
H_{S^1}
e^{\Phi \frac{\partial}{\partial \theta_1}}
= H_{S^1}(\Phi)$ for any $\Phi\in\RR$. 
Define the unitary operator
\begin{align*}
U(\Phi) = e^{\Phi \frac{\partial}{\partial \theta_1}}.
\end{align*}
Then the Josephson current is 
expressed as 
\begin{align}
(\psi, I_{S^1}(\Phi)\psi)=(U(\Phi)\psi, I_{S^1}(0)
U(\Phi)\psi ).
\end{align}
Let 
\begin{align*}
\psi(\theta_1,\theta_2)=
\sum_\pm\sum_{1\leq n\leq k-1} a_n^\pm e^{\pm in\theta_1} e^{i(k-n)\theta_2}+
a_0 e^{ik\theta_2}
\in L_{2k}.
\end{align*}
Then 
\begin{align*}
U(\Phi)
\psi
=
\sum_\pm\sum_{1\leq n\leq k-1} a_n^\pm e^{\pm in(\theta_1+\Phi)} e^{i(k-n)\theta_2}+
a_0 e^{ik\theta_2}.
\end{align*} 
Hence, one observes that 
\begin{align*}
\begin{array}{lll}
e^{\pm i n \theta_1} &\longrightarrow e^{\pm i n(\theta_1+\Phi)}& n\neq0,\\
1&\longrightarrow 1& n=0. 
\end{array}\end{align*}
Here, the index $\pm n$ represents the difference in the number of particles 
located in $\mathcal{H}_A$ and $\mathcal{H}_B$, respectively. 
For instance, the term $e^{i n \theta_1}$ corresponds to a configuration with 
$n+m$ particles in $\mathcal{H}_A$ and $m$ particles in $\mathcal{H}_B$ for any $m$. 
The situation may be interpreted as: 
\begin{align*}(n+m) \ \text{clockwise windings} + m \ \text{counterclockwise windings}.\end{align*}
Consequently, a phase shift $e^{i n \Phi}$ arises due to the Aharonov-Bohm effect.
%See Appendix \ref{abeffect} for the Aharonov-Bohm effect on the circle. 
Thus, the Josephson current in the presence of a magnetic field with respect to $\psi$ 
is equal to the Josephson current in the absence of a magnetic field with respect to the 
conjugated state $U(\Phi)\psi$, reflecting the Aharonov-Bohm effect.

\subsection{Fraunhofer pattern}
In the presence of a constant  magnetic field $B=(0,0,b)$ applied perpendicular to a Josephson junction, the Josephson current acquires a position-dependent  phase shift  along the width of the junction. 
Specifically, the vector potential $A$ induces a  phase shift  that varies linearly with the coordinate 
$x$ across the junction. 
As a consequence, the local Josephson current density oscillates as a function of 
$x$, and the total current flowing through the junction is obtained by integrating these contributions over the width of the device. 
This interference effect gives rise to the well-known Fraunhofer pattern, in which 
the critical current as a function of the  phase shift  through the junction exhibits the same envelope as the intensity distribution of single-slit diffraction in optics. 
The following computation provides a precise derivation of this Fraunhofer pattern.

We consider a Josephson junction with barrier thickness $d$ and width $W=1$. 
Let us consider a constant  magnetic field $B=(0,0,b)$, which is explained in 
Example \ref{constant }. 
Then the  phase shift  is given by 
\begin{align*}\Phi = \Phi(x)=\Psi x\qquad -1/2\leq x\leq 1/2,\end{align*} 
where $\Psi=bd$ is the magnetic flux. 
The Josephson current associated with $\Phi(x)$ is denoted by $I_{S^1}(\Phi(x))$. 
The total Josephson current associated with $\psi\in \cH_{S^1}$ is defined by
\begin{align*}
I_{\rm total}(\Psi)=\int_{-1/2}^{1/2} (\psi, I_{S^1}(\Phi(x))\psi) \rd x.
\end{align*}

\begin{theorem}[Fraunhofer pattern]
We have 
\begin{align*}
I_{\rm total}(\Psi)
= \aaa {\displaystyle \frac{\sin (\Psi/2)}{\Psi/2} (\psi, I_{S^1}(\Phi(0))\psi)& 
\psi \in \cH_{S^1}\setminus P_{\{0\}}\otimes P_{\{0\}}\cH_{S^1},\\
0& \psi \in P_{\{0\}}\otimes P_{\{0\}}\cH_{S^1}. 
}\end{align*}
\end{theorem}
\proof
\iffalse
 By the definition of $I_{S^1}(\Phi)$ we can compute 
 $I_{S^1}(\Phi)p_i\otimes p_j$. We set 
 $K_{ij}=I_{S^1}(\Phi)p_i\otimes p_j$ for the notational simplicity. We have 
\begin{align*}
\begin{array}{ll}
K_{11}=-2\sin(\theta_1-\Phi(x)-\theta_2)p_1\otimes p_1,  &
K_{21}=\lkk -ie^{-i(\theta_1-\Phi(x)-\theta_2)}+ie^{i(\theta_1-\Phi(x))}\rkk p_2\otimes p_1,\\
K_{31}=\lkk -ie^{-i(\theta_1-\Phi(x))}+ie^{i(\theta_1-\Phi(x)+\theta_2)}\rkk p_3\otimes p_1, &
K_{41}=-2\sin(\theta_1-\Phi(x)+\theta_2)p_4\otimes p_1, \\
K_{12}=ie^{i(\theta_1-\Phi(x)-\theta_2)}p_1\otimes p_2, &
K_{22}=ie^{i(\theta_1-\Phi(x))}p_2\otimes p_2, \\
K_{32}=ie^{-i(\theta_1-\Phi(x))}p_3\otimes p_2, &
K_{42}=-ie^{-i(\theta_1-\Phi(x)+\theta_2)}p_4\otimes p_2,\\
K_{13}=ie^{i(\theta_1-\Phi(x)+\theta_2)}p_1\otimes p_3, &
K_{23}=ie^{i(\theta_1-\Phi(x)+\theta_2)}p_2\otimes p_3,\\
K_{33}=0, &
K_{43}=-e^{-i(\theta_1-\Phi(x)-\theta_2)} p_4\otimes p_3,\\
K_{14}=-2\sin (\theta_1-\Phi(x)+\theta_2)p_1\otimes p_4,&
K_{24}=-2\sin (\theta_1-\Phi(x)+\theta_2)p_2\otimes p_4,\\
K_{34}=\lkk -ie^{-i(\theta_1-\Phi(x)+\theta_2)}+ie^{i(\theta_1-\Phi(x)-\theta_2)}\rkk p_3\otimes p_4,&
K_{44}=-2\sin (\theta_1-\Phi(x)-\theta_2)p_4\otimes p_4. 
\end{array}
\end{align*}
\fi
By the decomposition given by \kak{c7} we have 
\begin{align*}
\int_{-1/2}^{1/2} (\psi, I_{S^1}(\Phi(x))\psi) \rd x
=\sum_{1\leq i,j\leq 4}
\int_{-1/2}^{1/2} (p_i\otimes p_j \psi, I_{S^1}(\Phi(x))p_i\otimes p_j \psi) \rd x
\end{align*}
Let $\psi=p_4\otimes p_4 \psi$. 
We have
\begin{align*}
\int_{-1/2}^{1/2} (\psi, I_{S^1}(\Phi(x))\psi) \rd x
= 2\alpha \int_{-1/2}^{1/2} \rd x
\int_{S^1\times S^1} 
\overline{\psi(\theta_1,\theta_2)}
\sin(\theta_1-\Psi x-\theta_2)
\psi(\theta_1,\theta_2)
\rd\theta_1 \rd\theta_2.
\end{align*}
By the Fubini theorem, we can exchange the order of integration in $x$ and in $(\theta_1,\theta_2)$. 
Since 
\begin{align*}
\int_{-1/2}^{1/2} \sin(\theta_1-\Psi x-\theta_2) \rd x 
= \sin(\theta_1-\theta_2) \frac{\sin(\Psi/2)}{\Psi/2},
\end{align*}
we see that 
\begin{align}\label{fh1}
\int_{-1/2}^{1/2} (\psi, I_{S^1}(\Phi(x))\psi) \rd x
= \frac{\sin (\Psi/2)}{\Psi/2} (\psi, I_{S^1}(\Phi(0))\psi). 
\end{align}
Let $\psi=p_3\otimes p_2 \psi$. 
We have
\begin{align*}
\int_{-1/2}^{1/2} (\psi, I_{S^1}(\Phi(x))\psi) \rd x
= 2\alpha \int_{-1/2}^{1/2} \rd x
\int_{S^1\times S^1} 
\overline{\psi(\theta_1,\theta_2)}
ie^{-i(\theta_1-\Phi(x))}
\psi(\theta_1,\theta_2)
\rd\theta_1 \rd\theta_2.
\end{align*}
Since 
\begin{align*}
\int_{-1/2}^{1/2} ie^{-i(\theta_1-\Phi(x))}
\rd x 
= e^{-i\theta_1} \frac{\sin(\Psi/2)}{\Psi/2},
\end{align*}
we also have \kak{fh1}.  
Hence for $\psi$ such that $\psi=p_i\otimes p_j\psi$ for $(i,j)\neq (3,3)$,  
\kak{fh1} holds true. 
For $\psi=p_3\otimes p_3\psi$, 
\begin{align*}
\int_{-1/2}^{1/2} (\psi, I_{S^1}(\Phi(x))\psi) \rd x=0. 
\end{align*}
Then the proof is complete. 
\qed

\subsection{Vanishing of Fraunhofer pattern}
In this section, we present examples in which the Fraunhofer pattern vanishes. 
Let
\begin{align}
\label
{st5}
&\psi_0(\theta_1,\theta_2)=\sum_\pm\sum_{1\leq n\leq k-1} a_n^\pm e^{\pm in\theta_1} e^{i(k-n)\theta_2}+
a_0 e^{ik\theta_2}
\in L_{2k},\\
\label{st6}&\psi_1(\theta_1,\theta_2)=\sum_{0\leq n\leq k-2} 
(a_n^{+} e^{+ in\theta_1} +
a_n^{-} e^{- i(n+1)\theta_1})
e^{-i(k-n)\theta_2}
\in L_{2k-1}.
\end{align}
When $a_n^{+} = a_n^{-}$, we call $\psi_0$ a standing wave and $\psi_1$ a one-mode shifted standing wave
\begin{lemma}\label{st3}
Let $\psi=\psi_0$ be a standing wave. 
%Suppose that $a_n^{+}=a_n^{-}=a_n$ for all $n$. 
Then 
\begin{align}\label{c2}
(\psi, I_{S^1}(\Phi)\psi)=-8\pi \alpha C \sin\Phi, 
\end{align}
where $
C=2\pi \Re \sum_{0\leq n\leq k-2}
\bar a_{n+1} a_n$.
\end{lemma}
\proof
By \kak{c1} we can compute as 
$%\begin{align*}
(\psi, I_{S^1}(\Phi)\psi)
=
2\alpha
\int_{S^1\times S^1}
\sum_{j=1}^6 f_j \rd \theta_1\rd \theta_2$. 
%\end{align*}
The integrant consists of the six terms below:
\begin{align*}
\iffalse
&\sin(\theta_1-\Phi)
\sum_\pm\sum_{1\leq n'\leq k-1} \bar a_{n'}^\pm e^{\mp in'\theta_1} e^{-i(k-n')\theta_2}
 a_0e^{i(k-1)\theta_2}\\
&
+
\sin(\theta_1-\Phi+\theta_2)
\sum_\pm\sum_{1\leq n'\leq k-1} \bar a_{n'}^\pm e^{\mp in'\theta_1} e^{-i(k-n')\theta_2}
\sum_{1\leq n\leq k-1} 
a_n^{-} e^{-in\theta_1} e^{i(k-n)\theta_2}\\
&+
\sin(\theta_1-\Phi-\theta_2)
\sum_\pm\sum_{1\leq n'\leq k-1} \bar a_{n'}^\pm e^{\mp in'\theta_1} e^{-i(k-n')\theta_2}
\sum_{1\leq n\leq k-1} 
a_n^{+} e^{+in\theta_1} e^{i(k-n)\theta_2}\\
&+
\sin(\theta_1-\Phi)
\bar a_0 e^{-ik\theta_2}
 a_0e^{i(k-1)\theta_2}\\
&+
\sin(\theta_1-\Phi+\theta_2)\bar a_0 e^{-ik\theta_2}
\sum_{1\leq n\leq k-1} 
a_n^{-} e^{-in\theta_1} e^{i(k-n)\theta_2}\\
&+
\sin(\theta_1-\Phi-\theta_2)
\bar a_0 e^{-ik\theta_2}
\sum_{1\leq n\leq k-1} 
a_n^{+} e^{+in\theta_1} e^{i(k-n)\theta_2}\\
\fi
&
f_1=\sin(\theta_1-\Phi)
\sum_\pm\sum_{1\leq n'\leq k-1} \bar a_{n'}^\pm e^{\mp in'\theta_1} e^{i(n'-1)\theta_2}
 a_0,\\
&
f_2=
\sin(\theta_1-\Phi+\theta_2)
\sum_\pm\sum_{1\leq n',n\leq k-1} 
\bar a_{n'}^\pm e^{\mp in'\theta_1} 
a_n^{-} e^{-in\theta_1} e^{i(n'-n)\theta_2},\\
&f_3=
\sin(\theta_1-\Phi-\theta_2)
\sum_\pm\sum_{1\leq n',n\leq k-1} 
\bar a_{n'}^\pm e^{\mp in'\theta_1} e^{i(n'-n)\theta_2}
a_n^{+} e^{+in\theta_1}, \\
&f_4=
\sin(\theta_1-\Phi)
\bar a_0 
 a_0e^{-ik\theta_2},\\
&f_5=
\sin(\theta_1-\Phi+\theta_2)\bar a_0 
\sum_{1\leq n\leq k-1} 
a_n^{-} e^{-in\theta_1} e^{-in\theta_2},\\
&f_6=
\sin(\theta_1-\Phi-\theta_2)
\bar a_0 
\sum_{1\leq n\leq k-1} 
a_n^{+} e^{+in\theta_1} e^{-in\theta_2}. 
\end{align*}
Note that 
\begin{align}\label{sin}
\int_{S^1} \sin(\theta-\theta_2) e^{in\theta_2}\rd\theta_2=
\aaa{i\pi e^{-i\theta}& n=-1,\\
-i\pi e^{i\theta}& n=1,\\
0 &n\not=\pm 1}\end{align}
for any $n\in\ZZ$ and $\theta\in\RR$. 
By integrating above six terms 
on $S^1\times S^1$ 
by employing the formula \eqref{sin}, 
we can obtain 
\begin{align*}
\frac{1}{2\alpha}\frac{1}{2\pi}(\psi, I_{S^1}(\Phi)\psi)
&=
 i\pi e^{i\Phi}
\left( \bar a_{1}^{-}a_0+\sum_{1\leq n\leq k-2}
\bar a_{n+1}^{-} a_n^{-} 
+\bar a_0a_1^{+}+
\sum_{2\leq n\leq k-1}\bar a_{n-1}^{+} 
a_n^{+}\right)\\
&-i\pi e^{-i\Phi}
\left( \bar a_{1}^{+}a_0
+\sum_{1\leq n\leq k-2}
\bar a_{n+1}^{+} a_n^{+} 
+\bar a_0 a_1^{-}+\sum_{2\leq n\leq k-1}
\bar a_{n-1}^{-} a_n^{-} 
\right)\\
&= -2 C\sin \Phi. 
\end{align*}
Then \kak{c2} is proved. 
\qed
\begin{lemma}\label{st4}
Let $\psi=\psi_1$ be a one-mode shifted standing wave. 
%Suppose that $a_n^{+}=a_n^{-}=a_n$ for all $n$. 
Then 
\begin{align}\label{c4}
(\psi, I_{S^1}(\Phi)\psi)=-8\pi \alpha C \sin\Phi, 
\end{align}
where
$C=
2\Re \sum_{0\leq n\leq k-3}\bar a_{n+1}a_n$. 
\end{lemma}
\proof
Since
$(\psi, I_{S^1}(\Phi)\psi)
=
2\alpha\int_{S^1\times S^1}
(f_1+f_2)\rd \theta_1\rd \theta_2$, 
where 
\begin{align*}
&
f_1=
\sin(\theta_1-\Phi+\theta_2)\sum_{0\leq n',n\leq k-2} 
(\bar a_{n'}^{+} e^{-in'\theta_1} +\bar a_{n'}^{-} e^{i(n'+1)\theta_1})
 a_n^{+} 
 e^{i(n-n')\theta_2}
e^{+ in\theta_1}, 
\\&
f_2=
\sin(\theta_1-\Phi-\theta_2)\sum_{0\leq n',n\leq k-2} 
(\bar a_{n'}^{+} e^{-in'\theta_1} +\bar a_{n'}^{-} e^{i(n'+1)\theta_1})
 a_n^{-} e^{- i(n+1)\theta_1}
e^{i(n-n')\theta_2}, 
\end{align*}
we can see that 
\begin{align*}
\frac{1}{2\pi}
\frac{1}{2\alpha}(\psi, I_{S^1}(\Phi)\psi)
&=i\pi e^{i\Phi}\lk \sum_{1\leq n\leq k-2}\bar a_{n-1}^{+}a_n^{+}
+ \sum_{0\leq n\leq k-3}\bar a_{n+1}^{-}a_n^{-}
\rk\\
&
-i\pi e^{-i\Phi}\lk \sum_{1\leq n\leq k-2}\bar a_{n-1}^{-}a_n^{-}
 +\sum_{0\leq n\leq k-3}\bar a_{n+1}^{+}a_n^{+} 
\rk\\
&= -2 C\sin \Phi. 
\end{align*}
Then the proof is complete.
\qed
Let us consider a constant  magnetic field $B=(0,0,b)$. 
Then the  phase shift  is given by 
$\Phi = \Phi(x)=\Psi x$. 

\begin{theorem}[Vanishing of Fraunhofer pattern]
Let $\psi$ be a standing wave $\psi_0$ 
 or $\psi$ be a one-mode shifted standing wave $\psi_1$. 
Then for all $\Psi\in\RR$, 
\begin{align*}
I_{\rm total}(\Psi)=0.
\end{align*}
\end{theorem}
\proof
By Lemmas \ref{st3} and \ref{st4} we have
\begin{align*}
I_{\rm total}(\Psi)=-8\pi\alpha C\int_{-1/2}^{1/2}\sin (\Psi x)\rd x=0. 
\end{align*}
Then the theorem is proved. 
\qed

In the usual situation, the presence of a constant magnetic field induces a linear phase gradient 
$\Psi x$ along the width of the Josephson junction. 
The local Josephson current then interferes across the junction, giving rise to the characteristic Fraunhofer diffraction pattern.
However, on the standing wave state $\psi_0$ 
or the one-mode shifted standing wave state $\psi_1$
%associated with the number-phase relation, 
the current distribution becomes spatially uniform due to symmetry. This is shown in 
Lemmas~\ref{MAIN1} and \ref{MAIN2}.  
As a consequence, the spatial modulation that normally produces the Fraunhofer pattern 
is averaged out, and the interference fringes disappear. 
In other words, the current no longer carries information about the spatial phase shifts, 
and the total current becomes independent of the applied magnetic flux.

\section{Concluding remarks}\label{8}
From a mathematical standpoint, extending the study from a single Josephson junction 
to an array of $n$ junctions opens up new avenues in operator theory, e.g.,\cite{DeLuca2015}. 
The emergent higher-rank symmetries, such as the $\mathrm{SU}(3)$ symmetry that arises in the 
three-junction case, call for a rigorous investigation of the algebraic structures and spectral 
properties of the associated Hamiltonians. This direction promises to enrich the interplay 
between functional analysis and spectral theory, offering fresh 
insight into how symmetries are encoded in physically motivated operators.

On the physical side, Josephson junction networks provide a unique platform for realizing 
condensed matter analogues of phenomena usually associated with high-energy physics. 
The emergence of $\mathrm{SU}(3)$ symmetry in the $n=3$ case, echoing the structure of 
the strong interaction in the Standard Model, suggests a striking bridge between 
superconducting quantum devices and the symmetry principles underlying elementary particles. 
Such parallels indicate that Josephson networks may serve as experimental testbeds for 
exploring fundamental aspects of quantum field theory in a controlled laboratory setting.

\appendix
\section{Conjugate operators associated with $\sM$ and $\sN$}\label{conj}
The multiplication by $\theta$ is formally 
regarded as  a conjugate operator associated with
$-i \tfrac{\partial}{\partial \theta_1}$.
In Remark~\ref{np}, however,  
we pointed out that multiplication by $\theta_{1}$ is not a well-defined operator on 
$\mathcal{H}_{S^1}$.
Nevertheless, it can be shown that there exists a conjugate operator associated with
$-i \tfrac{\partial}{\partial \theta_1}$.
Let $f_n$ be the eigenvector of $\sM$ corresponding to the eigenvalue $n$. 
$T_G$  is defined by 
\[T_Gf=i\sum_{n=0}^\infty \sum_{m\neq n}\frac{(f_m,f)}{n-m}f_n,\]
as introduced in \cite{gal02a}. 
In \cite{HT24b} it is shown that $T_G$ can be represented in terms of shift operators 
$L$  and $L^\ast$ as 
\[T_G=i\lk \log(\one-L)+\log(\one-L^\ast)\rk.\]
Moreover 
one can regard ${\ln}^\ast\otimes \ln$ as the space of Hilbert-Schmidt operators on $\ln$. 
Under the identification ${\ln}^\ast\cong \ln$, we see that 
for $f\otimes g\in \ln\otimes \ln$, 
$(f\otimes g) (h)=(f,h)g$. Then $T_g$ can be also represented as 
\begin{align}\label{2ba}
T_G =i\sum_{n\neq m} \sN^{-1}(f_n\otimes f_m).
\end{align}
$T_G$ is a bounded self-adjoint operator on $\ln$ 
and it satisfies that 
\[[T_G, \sM]=-i\one \]
on $D=\ov{\rm LH} \{f_n-f_m\mid n,m\geq0\}$.   
Let us define 
\[\hat T_G=T_G\otimes \one-\one\otimes T_G\]
acting on $\cH$. 
Thus we  have 
\begin{align}
\label{conj1}
[\hat T_G, \sN]=-2i
\end{align}
on $D\otimes L^1(S^1)+ L^1(S^1)\otimes D$. 
Employing the unitary operator $\cU:\cH\to \cH_{S^1}$ we define 
\[\hat \theta_1=  \cU \hat T_G \cU^{-1}.\]
Form \kak{conj1} and 
$\cU \half \sN\cU^{-1}=-i\frac{\partial}{\partial \theta_1}$, 
the proposition below follows. 
\begin{proposition}[Conjugate of $-i\frac{\partial}{\partial \theta_1}$]
$\hat \theta_1$ is a bounded self-adjoint operator and it is a conjugate operator associated 
with $-i\frac{\partial}{\partial \theta_1}$: 
\[\left[\hat \theta_1, -i\frac{\partial}{\partial \theta_1}\right]=-i\one\]
on $\cU (D\otimes L^1(S^1)+ L^1(S^1)\otimes D)$. 
\end{proposition}

\section{Aharonov-Bohm effect}\label{app}
We refer the reader to \cite{ara92a,ara95a,ara95b} in this section. 
Let $R=\RR^2\setminus\{a_1,\dots,a_N\}$, and let
$A=(A_1,A_2)$ be a real-valued vector potential on $R$
with $A_j\in L^2_{\mathrm{loc}}(R)$, and let $q\in\RR$ denote the charge.
Set $\partial_1=\frac{\partial}{\partial x}$ and $\partial_2=\frac{\partial}{\partial y}$.
Define the symmetric operators
\[
P_j\psi=\bigl(-i\,\partial_j - q\,A_j\bigr)\psi,\quad j=1,2,
\]
with domain $D(P_j)= C_0^\infty(R)$. 
These are densely defined and closable, and we denote their closures by $\bar P_j$.  
For $(x,y)\in\RR^2$ and $s,t\in\RR$, let $C(x,y;s,t)$ be the
closed rectangle with base point $(x,y)$ and side lengths $|s|,|t|$, and let $D(x,y;s,t)$ denote its interior.
See Figure~\ref{cx}. 
\begin{figure}[h]
\centering
\begin{tikzpicture}[scale=1, every node/.style={font=\small}]
  % ---- parameters ----
  \def\x{1.0}   % lower-left x-coordinate
  \def\y{0.5}   % lower-left y-coordinate
  \def\s{3.0}   % width (s)
  \def\t{2.0}   % height (t)

  % ---- points ----
  \coordinate (P0) at (\x,\y);                 
  \coordinate (P1) at ({\x+\s},\y);            
  \coordinate (P2) at ({\x+\s},{\y+\t});       
  \coordinate (P3) at (\x,{\y+\t});            

  % ---- boundary with arrows (counterclockwise) ----
  \draw[thick,-{Latex[length=3mm]}] (P0) -- (P1);
  \draw[thick,-{Latex[length=3mm]}] (P1) -- (P2);
  \draw[thick,-{Latex[length=3mm]}] (P2) -- (P3);
  \draw[thick,-{Latex[length=3mm]}] (P3) -- (P0);

  % ---- labels ----
  \node[below left]  at (P0) {$\bigl(x,\,y\bigr)$};
  \node[below right] at (P1) {$\bigl(x+s,\,y\bigr)$};
  \node[above right] at (P2) {$\bigl(x+s,\,y+t\bigr)$};
  \node[above left]  at (P3) {$\bigl(x,\,y+t\bigr)$};

  % ---- interior marks ----
  \node (X1) at ({\x+0.8}, {\y+1.5}) {$\times$};
  \node[right=2mm] at (X1) {$a_{1}$};

  \node (X2) at ({\x+1.8}, {\y+0.9}) {$\times$};
  \node[right=2mm] at (X2) {$a_{2}$};
\end{tikzpicture}
\caption{$C(x,y;s,t)$}
\label{cx}
\end{figure}

\noindent
Let $B=\partial_1 A_2-\partial_2 A_1$ in $\mathcal{D}'(\RR^2)$, and define the magnetic flux by 
\[
\Phi_A(x,y;s,t)=\oint_{C(x,y;s,t)} A\cdot dr. 
\]

\begin{proposition}[{{\cite[Theorem 3.1]{ara95b}}}]
For all $s,t\in\RR$, the one-parameter unitary groups
$e^{is\bar P_1}$ and $e^{it\bar P_2}$ satisfy
\[
e^{is\bar P_1}\,e^{it\bar P_2} \;=\; 
e^{-iq\,\Phi_A(x,y;s,t)}\,e^{it\bar P_2}\,e^{is\bar P_1}.
\]
\end{proposition}

This relation encapsulates the Aharonov--Bohm effect: 
when the path winds once around the rectangle $C(x,y;s,t)$, 
the wave function acquires a phase shift given precisely by 
\[e^{-iq\,\Phi_A(x,y;s,t)}.\]
Let $Q_j$ denote multiplication by $x_j$. 
Then $[Q_i,Q_j]=0$. 
Moreover, $[P_i,P_j]=0$ if $B=~0$, and $[P_i,Q_j]=-i\delta_{ij}$. 
Thus $\{P_1,P_2, Q_1,Q_2\}$ furnishes a representation of the canonical commutation relations, though not necessarily equivalent to the Schr\"odinger representation 
$\{-i\partial_1,-i\partial_2,Q_1,Q_2\}$.  
We have the corollary below: 
\begin{corollary}[{{\cite[Corollary 3.4]{ara95b}}}]
 $\{P_1,P_2, Q_1,Q_2\}$ is equivalent to the Schr\"odinger representation 
$\{-i\partial_1,-i\partial_2,Q_1,Q_2\}$ if and only if
$\Phi_A(x,y;s,t)\in \dfrac{2\pi}{q}\,\mathbb{Z}$ for all $s,t\in\RR$ a.e.~$(x,y)$.
\end{corollary}

\subsection*{Acknowledgements}
FH is financially supported by JSPS KAKENHI 20K20886, JSPS KAKENHI 20H01808 and 
JSPS KAKENHI 25H00595.

\bibliographystyle{plain}
{\bibliography{hiro8}}

\end{document}